\documentclass[aps,prr,twocolumn,footinbib,showpacs,showkeys]{revtex4-2}

\usepackage{graphicx}
\usepackage{indentfirst}
\usepackage{physics}
\usepackage{braket}
\usepackage{float}
\usepackage{amsmath}
\usepackage{amssymb}
\usepackage{upgreek}
\usepackage{verbatim}
\usepackage{epstopdf}
\usepackage{CJK}
\usepackage{esint}
\usepackage{color}
\usepackage[T1]{fontenc}
\usepackage{subfigure}
\usepackage{amsfonts}
\usepackage{footmisc}
\usepackage{scrextend}
\usepackage{multirow}
\usepackage[hyperfootnotes=false]{hyperref}

\usepackage[english]{babel}
\usepackage{url}
\usepackage{bm}
\usepackage{hyperref}
\definecolor{darkblue}{rgb}{0,0,0.5}
\hypersetup{
colorlinks=true,
linkcolor=black,
filecolor=blue,
citecolor=darkblue,
urlcolor=black,
}

\newtheorem{theorem}{Theorem}

\newtheorem{corollary}[theorem]{Corollary}

\newtheorem{lemma}[theorem]{Lemma}

\newenvironment{proof}[1][Proof]{\noindent\textbf{#1.} }{\ \rule{0.5em}{0.5em}}

\newcommand{\calE}{{\cal E}}

\newcommand{\calI}{{\cal I}}

\newcommand{\calT}{{\cal T}}

\newcommand{\1}{^{(1)}}

\newcommand{\state}[1]{\ketbra{#1}{#1}}

\def\be{\begin{equation}}
\def\ee{\end{equation}}
\def\ba{\begin{eqnarray}}
\def\ea{\end{eqnarray}}
\usepackage{bm}

\newcommand{\QZ}[1]{{{\textcolor{black}{#1}}}}

\begin{document}

\title{Ultimate limits of approximate unambiguous discrimination} % 

\author{Quntao Zhuang$^{1,2}$}
\email{zhuangquntao@email.arizona.edu}
\affiliation{
$^1$Department of Electrical and Computer Engineering, University of Arizona, Tucson, AZ 85721, USA
\\
$^2$James C. Wyant College of Optical Sciences, University of Arizona, Tucson, AZ 85721, USA
}
\date{\today}

\begin{abstract}
Quantum hypothesis testing is an important tool for quantum information processing. Two main strategies have been widely adopted: in a minimum error discrimination strategy, the average error probability is minimized; while in an unambiguous discrimination strategy, an inconclusive decision (abstention) is allowed to vanish any possibility of errors when a conclusive result is obtained. In both scenarios, the testing between quantum states are relatively well-understood, for example, the ultimate limits of the performance are established decades ago; however, the testing between quantum channels is less understood. Although the ultimate limit of minimum error discrimination between channels has been explored recently, the corresponding limit of unambiguous discrimination is unknown. In this paper, we formulate an approximate unambiguous discrimination scenario, and derive the ultimate limits of the performance for both states and channels. In particular, in the channel case, our lower bound of the inconclusive probability holds for arbitrary adaptive sensing protocols. For the special class of `teleportation-covariant' channels, the lower bound is achievable with maximum entangled inputs and no adaptive strategy is necessary. 

\end{abstract}
%\keywords{Quantum Information, Quantum Physics, Optics.}

\maketitle

\section{Introduction}

%MED vs UD:
Quantum sensing~\cite{pirandola2018advances,Ruo,Degen,giovannetti2011advances} has enabled quantum advantages in various applications, such as positioning and timing~\cite{giovannetti2001quantum,zhuang2020entanglement}, target detection~\cite{Lloyd2008,tan2008quantum,barzanjeh2015,zhang2015,zhuang2017}, digital-memory reading~\cite{Qreading}, distributed sensing~\cite{zhuang2018distributed,proctor2018multiparameter,ge2018distributed,guo2020distributed,xia2019entangled,zhuang2020distributed}, entangled-assisted spectroscopy~\cite{Shi2020} and most prominently the Laser Interferometer Gravitational-wave Observatory (LIGO)~\cite{ligo,LIGO_2,tse2019_ligo}. \QZ{As fundamental tools for quantum sensing, various different strategies have been developed for quantum hypothesis testing in various scenarios.} In a minimum error discrimination (MED)~\cite{Helstrom_1967,Helstrom_1976,Yuen_1973,Holevo_1982} strategy, the overall error probability is minimized, while in an unambiguous discrimination (UD)~\cite{ivanovic1987differentiate,dieks1988overlap,peres1988differentiate,huttner1996,chefles1998unambiguous} strategy, an inconclusive result is allowed to vanish any possible errors. Both strategies have wide applications: MED is important for applications like target detection and digital-memory reading; and UD can be utilized in applications related to quantum key distribution protocols~\cite{duvsek2000unambiguous} and optimal cloning~\cite{duan1998probabilistic}. 
To make UD practically relevant, given that the experimentation of sensing protocols is never perfect, relaxations of exact UD have been considered in many different approaches, including allowing a fixed inconclusive probability~\cite{chefles1998strategies,zhang1999general,eldar2003mixed,fiuravsek2003optimal,nakahira2015generalized,touzel2007optimal,nakahira2016finding}, maximum-confidence~\cite{croke2006maximum,herzog2009discrimination}, error-margin tuning~\cite{hayashi2008state} and general cost-function approaches~\cite{nakahira2015finding,combes2015cost,nakahira2016finding}. 

Another complication beyond the different strategies is that the hypotheses being discriminated often involve physical processes modelled as quantum channels. While quantum state discrimination~\cite{Helstrom_1976,hirota1988properties,chefles2000quantum} is relatively well-understood, many open problems in quantum channel discrimination~\cite{KitaevDiamond,Acin_2001,sacchi2005entanglement,wang2006unambiguous,hayashi2016quantum} still await answers. For example, the ultimate limit of state MED~\cite{Helstrom_1967}, the general condition of UD~\cite{chefles1998unambiguous,feng2004unambiguous,wang2006unambiguous} and the optimum UD of various ensembles of states~\cite{chefles1998optimum,herzog2007optimum,pang2009optimum,kleinmann2010structural,bergou2012optimal,bandyopadhyay2014unambiguous} are known. However,
quantum channel discrimination is complicated by the various choices of input states and the potential of an adaptive strategy. Due to the complication, the ultimate limits of channel discrimination is much more challenging to solve. While the Helstrom limit~\cite{Helstrom_1967} is established half a century ago, the ultimate limit of MED between quantum channels has been unsolved until very recently~\cite{pirandola2019fundamental,zhuang2020ultimate}.
And the ultimate limit of UD between quantum channels, beyond the exact UD condition~\cite{wang2006unambiguous}, is not well-understood.

In this paper, we solve the ultimate limit of an approximate version of UD between quantum channels. Different from previous approaches, we enable the fine-tuning of all conclusive conditional error probabilities. Such a relaxation allows the proof of a general continuity inequality for states.
Then, we prove an ultimate lower bound on the inconclusive probability in approximate channel-UD for an arbitrary adaptive sensing protocol, following Refs.~\cite{pirandola2019fundamental,zhuang2020ultimate}. The lower bound can be calculated from the approximate UD between Choi states. For a special ensemble of channels called `jointly-teleportation-covariant' channels, this lower bound can be achieved by inputting the maximum entangled states and directly measuring the output, without any adaptive strategy involved. \QZ{When the Choi states are pure, this achievable lower bound can be directly calculated; while for mixed Choi states, we further obtain an efficiently-calculable but non-tight lower bound.}

%We begin our paper by introducing the approximate UD for states in Sec.~\ref{sec:states}, and then extend the approach to channels in Sec.~\ref{sec:channels}.

%\QZ{Ref.~\cite{croke2006maximum} here is worth reading.}

%\QZ{Most related paper are these~\cite{herzog2012optimized,sugimoto2012discrimination}, will need to read into them and those paper citing this one.}

\section{Approximate unambiguous discrimination between states}
\label{sec:states}

Consider an ensemble of states $\bm \Upsilon=\{\bm \rho, \bm P\}$, where the states $\bm \rho=\{\rho_n\}_{n=1}^m$ have prior probabilities $\bm P=\{P_n\}_{n=1}^m$. The goal of a state discrimination task is to determine an unknown state sampled from the ensemble through a measurement. In general, we also allow an inconclusive decision, when not enough information is obtained to reach a definitive conclusion. Therefore, the measurement is described by a set of POVMs $\bm \Pi=\{\Pi_n\}_{n=0}^m$, where $\Pi_0$ represents the inconclusive result, and each of the rest $\Pi_n$ corresponds to the decision that the state is $\rho_n$. As POVMs, we require each element $\Pi_n\ge0$ to be positive semi-definite, and $
\sum_{n=0}^m \Pi_n=I
$,
with $I$ being the identity operator.

%To characterize the performance, we denote the conditional probability $P_{n^\prime|n}=\tr\left(\rho_n \Pi_{n^\prime}\right)$; therefore the conditional inconclusive probability and correct probability are given by
%$ 
%P_{0|n}
%$
%and 
%$
%P_{n|n}
%$ 
%respectively. 
The performance of the protocol is characterized by an overall inconclusive probability
\be 
P_F=\sum_n P_n \tr\left(\rho_n \Pi_{0}\right),
\label{PF_def}
\ee 
where `F' stands for `failure', and the conditional conclusive error probabilities
\be 
P_{E|n}=1-\left[\tr\left(\rho_n \Pi_{n}\right)+\tr\left(\rho_n \Pi_{0}\right)\right],
\label{PEn_def}
\ee 
where `E' stands for `error'.

In the exact UD scenario, one requires all $P_{E|n}$ to be zero; in this paper, we introduce the approximate version of UD, where each of the conclusive error probability $P_{E|n}$ is bounded, forming the set of constraints
\be 
\mathbb{E}_U\left(\bm \epsilon\right)=\{P_{E|n}\le \epsilon_n\}_{n=1}^m,
\label{epsilon_constraints}
\ee 
where $\bm \epsilon=\{\epsilon_n\}_{n=1}^m$ describes the error tolerance. Alternatively, we can also consider the error probability conditioned on making a conclusive decision. Namely, the constraints become 
\be 
\mathbb{E}_R\left(\bm \epsilon\right)=\{\frac{P_{E|n}}{1-P_F}\le \epsilon_n\}_{n=1}^m.
\label{epsilon_constraints_R}
\ee

Under the above constraint, in general we want to choose the POVM $\bm \Pi$ to minimize the inconclusive probability to obtain
\be 
P_F^{X\star}(\bm \epsilon;\bm \Upsilon)=\min_{\bm \Pi:\mathbb{E}_X\left(\bm \epsilon\right)} P_F(\bm \Upsilon, \bm \Pi)
\label{PF_star_def}
\ee 
as a function of the state ensemble $\bm \Upsilon$ and the constants $\bm \epsilon$. Here `X' is a superscript to denote the different constraints: when `$X=U$' we adopt the un-rescaled constriants in Eq.~\eqref{epsilon_constraints}; when `$X=R$' we adopt the rescaled constraints in Eq.~\eqref{epsilon_constraints_R}.

The different constraints can be adopted for different purposes. The constraints in Eq.~\eqref{epsilon_constraints_R} vary quickly when $P_F$ is close to unity, therefore leading to complexity in terms of obtaining a good continuity bound; Moreover, as we will show in Lemma~\ref{lemma:convexity}, $P_F^{U\star}$ is convex in the constraints $\bm \epsilon$, while $P_F^{R\star}$ is not. However, $P_F^{R\star}$ is often easier to evaluate. 

%Previous works mainly concern the overall average conclusive error probability or the error probability conditioned on each {\em measurement result}, our approach differs to enable the fine tuning of each conditional error probability, which enables our results in the channel case.

As the constraints are closely connected by a rescaling, we expect a close connection between $P_F^{U\star}$ and $P_F^{R\star}$. Ref.~\cite{fiuravsek2003optimal} minimize the overall error probability given a fixed $P_F$, and two constrains are indeed equivalent there. In our case, the formulation is slightly more involved. We can show that when $P_F^{U\star}(\bm \epsilon;\bm \Upsilon)$ as a function of $\bm \epsilon$ is strictly convex or $P_F^{U\star}(\bm 0;\bm \Upsilon)<1$ then we have (proof in Appendix)
\be 
P_F^{U\star}(\bm \epsilon^U;\bm \Upsilon)=P_F^{R\star}(\bm \epsilon^R=\frac{1}{1-P_F^{U\star}(\bm \epsilon^U;\bm \Upsilon)}\bm \epsilon^U;\bm \Upsilon ).
\label{PF_connections}
\ee 
Here we utilize superscripts `U' or `R' in the dummy variables $\bm \epsilon$ in the constraints for clarity.

Our formulation allows an interpolation between UD and MED. 
The case of $\bm \epsilon=\bm 0$ in both cases corresponds to the (exact) UD scenario. When exact UD is possible, we have $P_F^{X\star}\left(\bm 0;\bm \Upsilon\right)<1$; while if exact UD is not possible, we have $P_F^{X\star}\left(\bm 0;\bm \Upsilon\right)=1$, which can be achieved by answering `inconclusive' all the time ($\Pi_0=I$); By fixing the \QZ{inconclusive probability} to be zero, one can obtain the Helstrom limit
\be
P_H(\bm \Upsilon)=\min_{\bm \epsilon:P_F^{X\star}(\bm \epsilon;\bm \Upsilon)=0} \bm P\cdot \bm \epsilon,
\label{Helstrom}
\ee 
therefore recover the MED case. 

To illustrate the scenario, we consider a pair of non-identical pure states $\{\ket{p},\ket{q}\}$, with a general prior $\{p,q\}$ and overlap $\braket{p|q}=\xi>0$. We denote this ensemble $\bm \Upsilon_{p,q}$ for simplicity. Following the ancilla-assisted measurement protocol~\cite{dieks1988overlap,peres1988differentiate,jaeger1995optimal}, we can solve the rescaled problem with constraint $\{\epsilon_p,\epsilon_q\}$ and obtain (see Appendix~\ref{sec:binary_pure_state_details} for details) the inconclusive probability 
\begin{align}
&P_F^{R\star}=g(\xi;\epsilon_p,\epsilon_q)
\equiv\min_{\tilde{\delta},\tilde{\beta}} p \sin^2 \tilde{\beta}+q \sin^2 \tilde{\delta}, 
\nonumber
\\
&\tilde{\delta},\tilde{\beta}:
\left(\xi-\sin \tilde{\beta}\sin \tilde{\delta}\right)/\left(\cos\tilde{\beta}\cos\tilde{\delta}\right)\in [\epsilon_{pq-},\epsilon_{pq+}],
\label{PF_opt_main_g}
\end{align}
where $
\epsilon_{pq\pm}=
|\sqrt{\epsilon_p(1-\epsilon_q)}\pm \sqrt{\epsilon_q(1-\epsilon_p)}|$.
One can also obtain a lower bound
\be 
P_F^{R\star}\ge h\left(\xi;\epsilon_p,\epsilon_q\right)\equiv 2\sqrt{pq}\left(1-\frac{1-\xi}{1-\epsilon_{pq+}}\right).
\label{PF_lower_bonund_h}
\ee 
This lower bound is achievable when $p=q=1/2$, from which we can see that convexity does not hold for the rescaled inconclusive probability. 

Because the ensemble of two non-identical pure states allows exact UD, we have $P_F^{X\star}\left(\bm 0;\bm \Upsilon_{p,q}\right)<1$ and therefore Eq.~\eqref{PF_connections} holds true and we can obtain the un-rescaled inconclusive probability from 
\be 
P_F^{U\star}\left[\{\epsilon_p^U,\epsilon_q^U\};\bm \Upsilon_{p,q}\right]=
g\left(\xi;\epsilon_p,\epsilon_q\right),
\label{PF_connections_g}
\ee 
where the constants $\epsilon_{p,q}^U=\left(1-g\left(\xi;\epsilon_p,\epsilon_q\right)\right)\epsilon_{p,q}$.
One can consider $\left\{\epsilon_p,\epsilon_q\right\}$ as the parameterization and obtain the overall curve of $P_F^{U\star}$.

\begin{figure}
\centering
\includegraphics[width=0.475\textwidth]{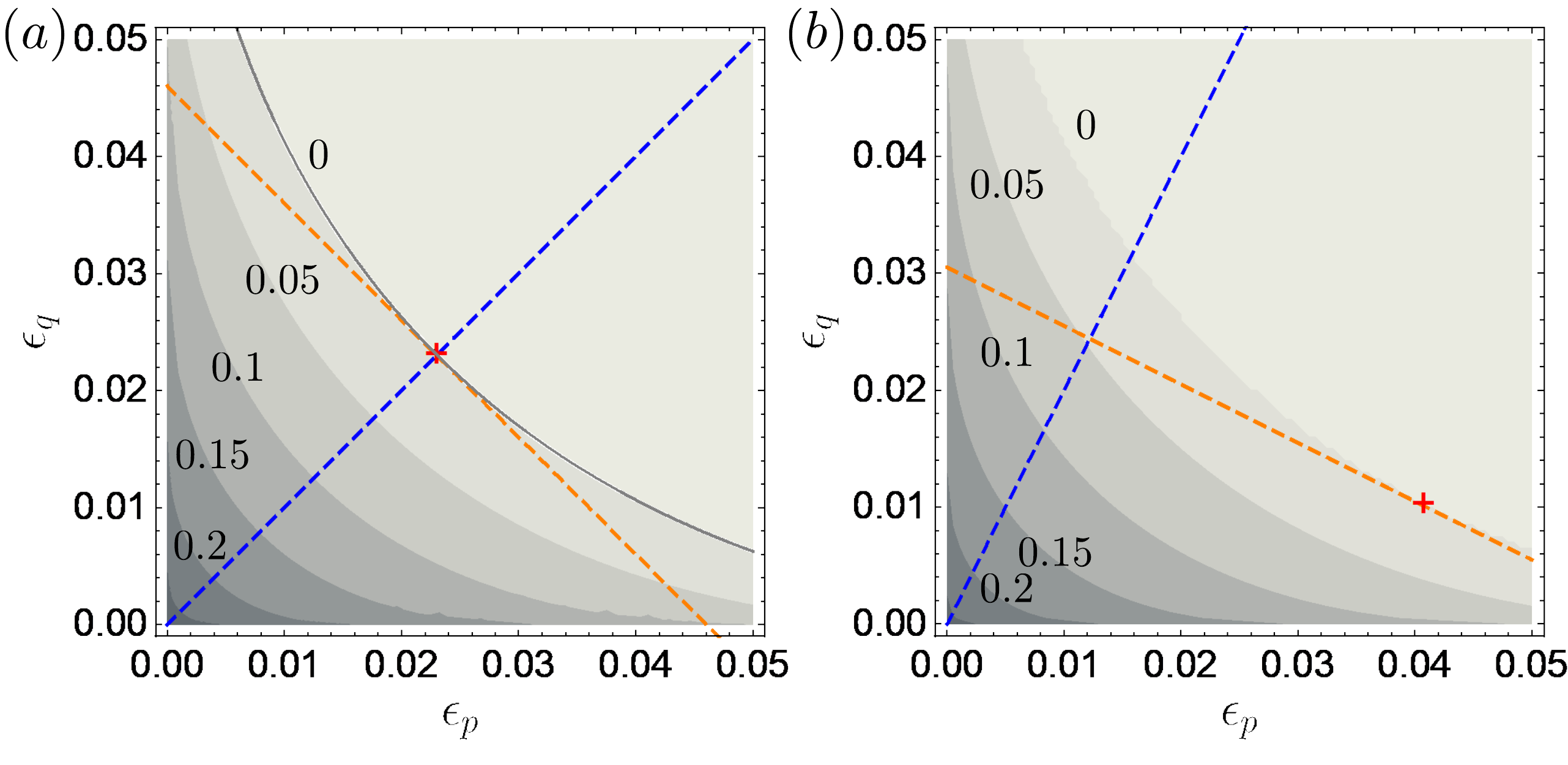}
\caption{(a) Contour of the asymmetric case Eq.~\eqref{PF_lower_bonund_h} of prior $p=q=1/2$ and overlap $\xi=0.3$. The red cross indicates the Helstrom limit is achieved. The blue dashed line is $\epsilon_q=\epsilon_p$. (b) Contour of the asymmetric case Eq.~\eqref{PF_opt_main_g} of prior $p=1/3,q=2/3$ and overlap $\xi=0.3$. The red cross indicates the Helstrom limit is achieved. The blue dashed line represents the vector $\bm P=(1/3,2/3)$. The orange dashed line is a tangent line to the $P_F^{U\star}=0$ boundary.
\label{fig:PF_contour}
}
\end{figure}

In Fig.~\ref{fig:PF_contour}, we plot the contour of the minimum \QZ{inconclusive probability} $P_F^{R\star}$ with $\epsilon_p,\epsilon_q$. In subplot (a), we utilize Eq.~\eqref{PF_lower_bonund_h} for the equal prior case. In subplot (b), we numerically solve Eq.~\eqref{PF_opt_main_g} for a asymmetric case of $p=1/3$ and $q=2/3$. Points with an equal error probability, $\bm P\cdot \bm \epsilon={\rm constant}$, lie on a surface orthogonal to the vector $\bm P$ (blue dashed). Since we require $P_F=0$, the minimum $\bm P\cdot \bm \epsilon$ is achieved when the equal error probability line is tangent to the region of $P_F=0$, as shown by the orange dashed line. The optimum choice of $\bm \epsilon$ lies at the tangent point (red cross).
%In our formulation of approximate UD, different from previous approaches, we extend the interpolation between UD and MED to an entire region. 

%In particular, we can verify that the solution obtained in Ref.~\cite{chefles1998strategies} is the cross section of the region by fixing equal prior and equal conditional error probabilities. And Ref.~\cite{zhang1999general} solved the equal prior case for the binary state, without fully exploring the error probability space.  

%\subsection{Convexity, continuity and data processing}
Now we derive properties of the constrained minimum inconclusive probability.

\subsection{Convexity}
First, we show the convexity of the inconclusive probability as a function of the error constraints.
\begin{lemma}
\label{lemma:convexity}
Convexity: consider constants $\{\bm \epsilon^{(k)}\}_{k=1}^K$ and non-negative numbers $\{r_k\}_{k=1}^K$ such that $\sum_{k=1}^K r_k=1$, for any ensemble of states $\bm \Upsilon=\{\bm \rho,\bm P\}$ we have
\be 
P_F^{U\star}(\sum_{k=1}^K r_k \bm \epsilon^{(k)};\bm \Upsilon)\le \sum_{k=1}^K r_k P_F^{U\star}(\bm \epsilon^{(k)};\bm \Upsilon).
\label{PF_ineq}
\ee 
\end{lemma}
The proof of this lemma is via constructing a mixed strategies to achieve the performance on the right side of Ineq.~\eqref{PF_ineq}~(see Appendix). Note that for the definition $P_F^{R\star}$ in Eq.~\eqref{PF_star_def}, convexity is not true, due to the extra factor in the constraint in Eq.~\eqref{epsilon_constraints_R}.

Convexity can be useful for obtaining bounds on the performance by extending some operation points.
We have the end point $P_F^{U\star}(0;\bm \Upsilon)\le 1$ as we explained, and $P_F^{U\star}(\bar{\bm P};\bm \Upsilon)=0$ from the random guessing strategy. Here $\bar{\bm P}=\{1-P_k\}_{k=1}^m$ is the complement of the prior. Applying Ineq.~\eqref{PF_ineq} for $K=2$, we have an upper bound on the binary state optimum inconclusive probability. For example, for a symmetric prior, we have
$
P_F^{U\star}(\{\epsilon\};\bm \Upsilon)\le 1- \epsilon m/ (m-1).
$
Better bounds can be achieved by considering the convex-hull of $P_F^{U\star}(0;\bm \Upsilon)\le 1$ and the point achieving the Helstrom limit.

\subsection{A data-processing inequality and lower bounds}
\label{sec:data_processing}
Similar to other sensing scenarios, we can obtain a data-processing inequality. Formally, for any quantum channel $\Lambda$, we have 
\be
P_F^{X\star}\left(\bm \epsilon^X;\{\Lambda\left(\bm \rho\right),\bm P\}\right) \ge P_F^{X\star}\left(\bm \epsilon^X;\{\bm \rho,\bm P\}\right),
\label{Eq:data_processing}
\ee 
for both $X=U,R$, as channel $\Lambda$ can always be applied on states first if one chooses to. 

%\QZ{Note that better bounds may be possible if one further consider the decreasing of the conditional errors when the states $\bm \rho$ are provided instead of $\Lambda\left(\bm \rho\right)$, we will defer that to future work.}

Using this data-processing inequality, we can obtain several lower bounds on the inconclusive probability. Consider mixed states $\bm \rho$ with purifications $\bm \psi_{\bm \rho}=\{\psi_{\rho_n}\}_{n=1}^m$, because $\bm \rho$ can be obtained by getting rid of the ancilla that purifies each state, Ineq.~\eqref{Eq:data_processing} leads to
\be 
P_F^{X\star}\left(\bm \epsilon^X;\{\bm \rho,\bm P\}\right) \ge P_F^{X\star}\left(\bm \epsilon^X;\{\bm \psi_{\bm \rho},\bm P\}\right),
\ee 
for both $X=U,R$. Utilizing Uhlmann's theorem that the maximum overlap between purification equals the fidelity $|\braket{\psi_{\rho_p}|\psi_{\rho_q}}|\le F\left(\rho_p,\rho_q\right)\equiv \tr \left(\sqrt{\sqrt{\rho_q} \rho_p \sqrt{\rho_q}}\right)$, we have the following lemma (see Appendix for proof).
\begin{lemma}
\label{lemma:fidelity_LB}
Consider a pair of general nonidentical quantum states $\{\rho_p,\rho_q\}$ with prior $\{p,q\}$, the minimum failure probability as a function of the rescaled tolerance $\{\epsilon_p^R,\epsilon_q^R\}$ satisfies
\be 
P_{F}^{R \star}\left(\{\epsilon_p^R,\epsilon_q^R\};\bm \Upsilon\right)\ge P_{F,LB,1}^{R \star}\equiv  g(F\left(\rho_p,\rho_q\right);\epsilon_p^R,\epsilon_q^R).
\label{PF_lower_bonund_exact}
\ee 
Equality is achieved when both states $\rho_p,\rho_q$ are pure.

A further analytical lower bound can be obtained
\be 
P_{F}^{R \star}\left(\{\epsilon_p^R,\epsilon_q^R\};\bm \Upsilon\right)\ge P_{F,LB,2}^{R\star}=h\left(F\left(\rho_p,\rho_q\right);\epsilon_p^R,\epsilon_q^R\right).
\label{PF_lower_bonund_anal}
\ee 
We have $P_{F,LB,1}^{R \star}\ge P_{F,LB,2}^{R\star}$, equal when $p=q=1/2$. 
\end{lemma}

The above lemma also allows lower bounds for $P_F^{U\star}$. Because the bounds are obtained from the purification, Eq.~\eqref{PF_connections} holds. Therefore, one can numerically obtain lower bounds by keeping track of the rescaling of the constraints.
%However, in general the relation in Eq.~\eqref{PF_connections} does not allow explicit expression of $\bm \epsilon^R$ as a function of $\bm \epsilon^U$, due to the fact that Eq.~\eqref{PF_opt_main_g} does not have a closed-form solution in general. Even for the case of $p=q=1/2$, utilizing  Eq.~\eqref{PF_lower_bonund_h}, $\bm \epsilon^R$ as a function of $\bm \epsilon^U$ is too lengthy to be practically useful.
We will apply this lower bound on two ensembles of mixed states that are relevant for our later analyses of the channel case. To narrow down the ultimate performance, we also provide upper bounds on the minimum inconclusive probability by designing explicit measurement strategies.

%\subsubsection{Binary mixed states with depolarizing noise}
%\label{sec:binary_mixed_depo}

\begin{figure}
\centering
\includegraphics[width=0.475\textwidth]{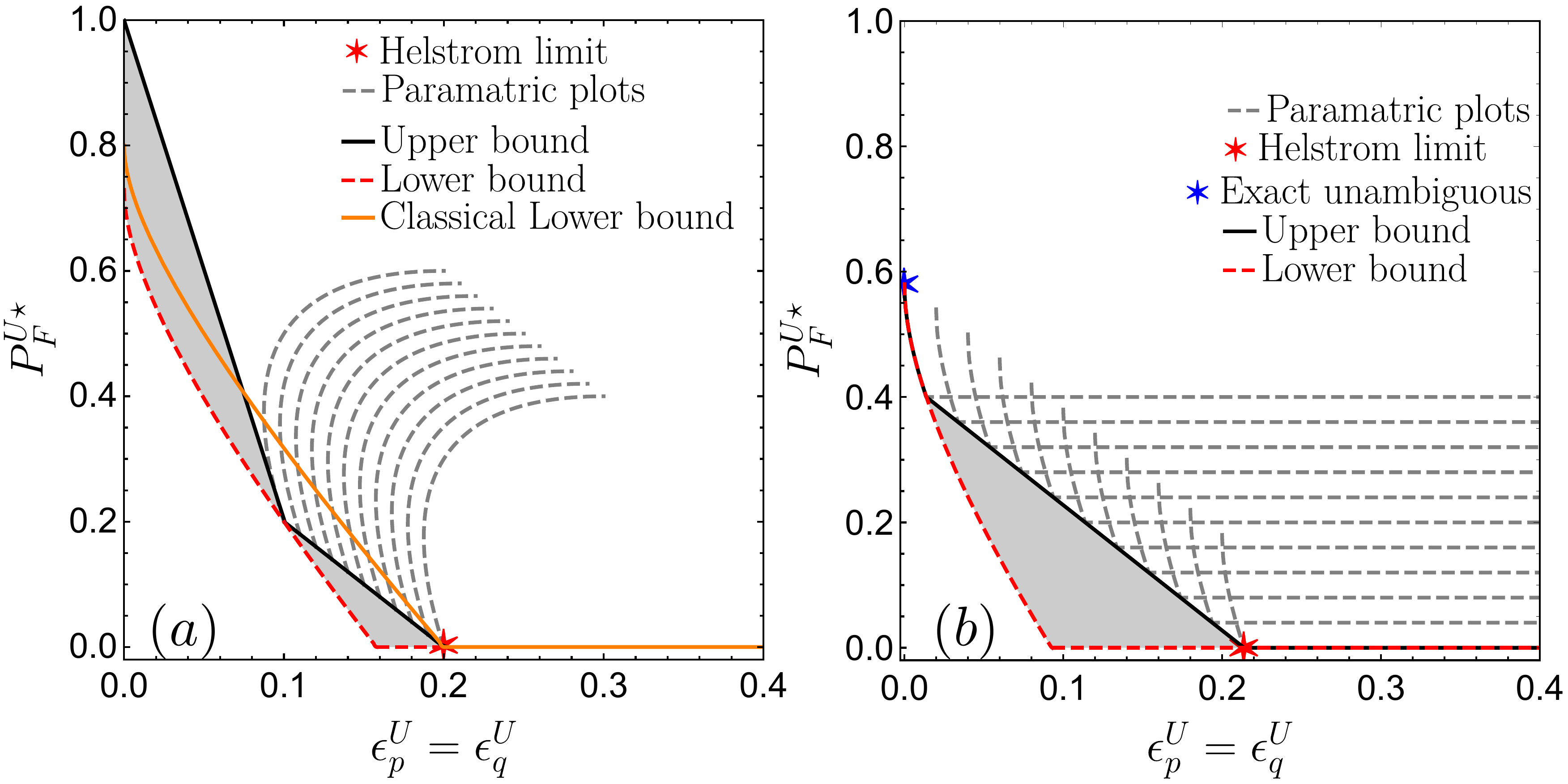}
\caption{ Inconclusive probability $P_F^{U\star}$ for UD between binary mixed states, with equal prior $p=q=1/2$ and symmetric error probability constraint $\epsilon_p^U=\epsilon_q^U$. The actual value is in the gray region between the lower bound (red solid, Eq.~\eqref{PF_lower_bonund_anal}) and the upper bound (black solid). The Helstrom limit is highlighted with the red star. We choose the parameter  $\eta=0.6$ and $\braket{p|q}=0.3$ in both cases. (a) Mixed states with depolarizing noise. The upper bound is obtained from the convex-hull of $(0,1)$, and the minimum over a family of strategies parameterized by $a$. \QZ{The classical lower bound is given in Eq.~\eqref{LB_classical}.} (b) Mixed states with a single common component. The upper bound and lower bound overlap when the error tolerance is small. The exact UD with inconclusive probability $\eta \braket{p|q}+\left(1-\eta\right)$ is marked in blue star.
In both (a) and (b), the strategies can achieve the dashed gray curves for different $a=0,0.1,\cdots,0.9,1$ from left to right.
\label{fig:PF_epsilon}
}
\end{figure}

{\em Binary mixed states with depolarizing noise.---}
We consider the approximate UD between an ensemble of two-qubit mixed states of equal priors,
\be 
\rho_{\pm}=\eta \psi_{\pm}+\frac{1-\eta}{4}I_4,
\label{rho_pm}
\ee 
where the pure states  $\ket{\psi_\pm}=\left(\ket{00}\pm\ket{11}\right)/\sqrt{2}$ are orthogonal maximal entangled states. This example of two-qubit states will also be utilized in the extension to channels in Sec.~\ref{sec:channel_example}.
The fidelity between states $\rho_\pm$ can be calculated analytically as
\begin{align}
&F\left(\rho_+,\rho_-\right)=\frac{1}{2}\left(1-\eta+\sqrt{1+2\eta-3\eta^2}\right),
\label{fideility_example_1}
\end{align}
and then the lower bound Eq.~\eqref{PF_lower_bonund_anal} can be analytically obtained for any error probability constraints. As shown in Fig.~\ref{fig:PF_epsilon}, the lower bound (red dashed) is below unity at the exact UD limit of zero error and goes to zero earlier than the Helstrom limit point (red star). This is because the bound is in general non-tight.

To provide an upper bound, we obtain the optimum strategy among a restricted class of strategies $\bm \Pi_{a,\theta}$ parameterized by $a,\theta$. Noticing the structure of the states, we design a measurement with POVMs
$
\Pi_Q=\state{\psi_+}+\state{\psi_-}
$ 
and $\Pi_P=I_4-\Pi_Q$
to separate them. If the result is $\Pi_P$, with probability $a\in[0,1]$ we decide inconclusive, with probability $1-a$ we perform random guess according to the prior; If the result is $\Pi_Q$, similar to Ref.~\cite{fiuravsek2003optimal} we consider the following POVMs parameterized by $\theta\in[\pi/2,\pi]$ as the second step
\begin{align}
&\Pi_0=\left(1-\frac{1}{\tan^2\left(\theta/2\right)}\right) \state{00},
\\
&\Pi_{\pm}=\frac{1}{2\sin^2(\theta/2)} \state{\phi_{\pm\theta}},
\end{align}
with 
$ 
\ket{\phi_\theta}=\cos\left(\theta/2\right)\ket{00}+ \sin \left(\theta/2\right)\ket{11}
$. The overall inconclusive probability and error probability are
\begin{align}
&P_F=\frac{a\left(1-\eta\right)}{2}+\frac{1+\eta}{4}\left(1-\frac{1}{\tan^2\left(\theta/2\right)}\right),
\\
&P_E=\frac{\left(1-\eta\right)\left(1-a\right)}{4}+\frac{1}{4\sin^2(\theta/2)}\left[\frac{1+\eta}{2}-\eta\sin\left(\theta\right)\right].
\end{align} 
One can then obtain an upper bound of $P_F^{U^\star}$ through convex-hull of the above strategy and $(\bm 0,1)$, as shown by the black solid line in Fig.~\ref{fig:PF_epsilon}(a). It is worthy to point out that the strategies above can also achieve the Hellstrom limit $P_H=\left(1-\eta\right)/{2}$ (shown by the red star). 

%\subsubsection{Binary mixed state with a single common component}
%\label{sec:binary_mixed_common}
{\em Binary mixed states with erasure.---}
Consider a noise model, where two pure states in $\bm \Upsilon_{p,q}$ are mixed with a common pure  orthogonal state $\phi$, via
\be 
\rho_x=\eta \state{x}+(1-\eta)\phi,
\label{states_example_2}
\ee 
for $x=p,q$.
We can evaluate the lower bound in Lemma~\ref{lemma:fidelity_LB} from the fidelity 
\be 
F=\eta |\braket{q|p}|+(1-\eta)=\eta \xi+(1-\eta).
\label{fideility_example_2}
\ee
As shown by the red dashed line in Fig.~\ref{fig:PF_epsilon}(b), the lower bound achieves the exact UD for the zero-error case, however, becomes zero earlier than the Helstrom limit (red star), which can be evaluated numerically.

%, for equal prior the expression $P_H=\left(1-\eta \sqrt{1-|\braket{p|q}|^2}\right)/2$. 

To obtain an upper bound, we design a two-step measurement. First, we perform the POVM $\{\phi,I-\phi\}$. If the outcome is $\phi$, we conclude inconclusive with probability $a\in[0,1]$, and randomly guess according to the prior with probability $1-a$; otherwise we proceed with the optimum strategy for the binary pure states case. The inconclusive probability and error constraints are given by
\begin{align}
&P_F^U 
=
\eta P_F^{U\star} \left(\{\epsilon_p^{U\prime},\epsilon_q^{U\prime}\};\bm \Upsilon_{p,q}\right)+(1-\eta)a,
\\
&\epsilon_p^{U}=(1-\eta)(1-a)q+\eta \epsilon_p^{U\prime},
\\
&\epsilon_q^{U}=(1-\eta)(1-a)p+\eta \epsilon_q^{U\prime},
\end{align}
where the function $P_F^{U\star}\left(\cdot;\bm \Upsilon_{p,q}\right)$ is given by Eq.~\eqref{PF_connections_g}.
By evaluating the above for different values of $a$ and $\epsilon_p^{U\prime},\epsilon_q^{U\prime}$, we can obtain the upper bound $P_F^U$ as a function of the overall error probabilities $\epsilon_p^{U},\epsilon_q^{U}$, as shown by the black solid line in Fig.~\ref{fig:PF_epsilon}(b).

%\begin{lemma}
%\label{lemma:fidelity_LB}
%Consider binary mixed state $\{\rho_p,\rho_q\}$ with prior $\{p,q\}$, the minimum failure probability as a function of the tolerance $\{\epsilon_p,\epsilon_q\}$ is lower bounded by
%\be 
%P_{F,LB}^\star=2\sqrt{pq}\left(1-\frac{1-F\left(\rho_p,\rho_q\right)}{1-\epsilon_{pq+}}\right),
%\label{PF_lower_bonund}
%\ee 
%where $
%\epsilon_{pq+}=
%|\sqrt{\epsilon_p(1-\epsilon_q)}+ \sqrt{\epsilon_q(1-\epsilon_p)}|$. It is achievable when the states are pure, and $p=q=1/2$.

\subsection{Continuity}

Continuity is an important property for a physical quantifier. Quantum states and channels are theory models of physical processes. As accurate as it can be, a theoretical description has unavoidable deviations from the reality. In this scenario, a continuous quantifier robust against imperfections is desired. While exact UD is theoretically well-studied, however, in practice when an ensemble of states is affected by a tiny bit of depolarizing noise, some ambiguity in the discrimination is unavoidable. While the property of exact UD is not continuous in quantum states, the relaxation to an approximate UD scenario allows a continuity bound (see Appendix for a proof).
\begin{lemma}
\label{approximate_continuity}
{\it Continuity of approximate UD:} consider two set of states $\bm \rho=\{\rho_n\}_{n=1}^m$ and $\bm \rho^\prime=\{\rho_n^\prime\}_{n=1}^m$, with one-norm deviation
$
\|\rho_n-\rho_n^\prime\|\le \delta_n, 1\le n \le m.
$
Given identical prior $\bm P=\{P_n\}_{n=1}^m$, the minimum failure probability $P_F^{U\star}(\bm \epsilon; \{\bm \rho,\bm P\})$ as a function of the un-rescaled tolerance $\bm \epsilon$
satisfies the continuity
\begin{align}
&P_F^{U\star}\left(\bm \epsilon;\{\bm \rho,\bm P\}\right)
\in [P_-,P_+], \mbox{ with}
\\
&P_\pm=P_F^{U\star}(\bm \epsilon\mp \bm \delta; \{\bm \rho^\prime,\bm P\})\pm \frac{1}{2}\bm P\cdot \bm \delta
\label{Eq:continuity}
\end{align}
where we denote $\bm \delta=\{\delta_k\}_{k=1}^n$.
\end{lemma}
Here, we focus on the un-rescaled constraints. For rescaled constraints, the continuity lower bound is more complicated and thus we discuss it in Appendix.

%\subsubsection{Cyclic pure states}
%Ref.~\cite{chefles1998optimum} solved the optimum exact UD for this case, with explicit measurement called equal-probability measurement (EPM). Consider an ensemble of pure states with GUS~\cite{cariolaro2010theory}, i.e., it has equal priors $P_n=1/m$ and the states satisfy
%\be 
%\phi_n=S^{(n-1)} \phi_1 S^{\dagger (n-1)}, 1\le n \le m
%\ee 
%where the unitary $S^m$ equals identity.

\section{Approximate unambiguous discrimination between channels}
\label{sec:channels}
Now we proceed to address the approximate UD between quantum channels.
In a channel discrimination scenario, one aims to perform hypothesis testing between a set of channels $\bm \calE=\{\calE_n\}_{n=1}^{m}$, with prior probabilities $\bm P=\{p_n\}_{n=1}^{m}$. 
To do that, one inputs quantum states and measure the output. In general, one can adopt an entanglement-assisted adaptive protocol $\mathbb{P}_u$ of $u$ channel uses~\cite{zhuang2020ultimate,pirandola2019fundamental}, where each operation can depend on all previous measurement results at earlier times and unlimited entanglement can be utilized. In this protocol, one can access an unknown channel $\calE$ for $u\ge 1$ times, as shown Fig.~\ref{fig:schematic_general}(a). With unlimited entanglement and unlimited ancillary systems allowed, all measurements can be pushed to the final output $\rho_{\calE,u}$, by introducing controlled unitaries. In each round of probing, a subsystem $S_k, 1\le k \le u$, in an arbitrary quantum state, is sent through the channel $\calE$ and the output $S_k^\prime$ is collected for later use. In an adaptive strategy, the quantum state of the probe $S_{k+1}$ in the next round can be produced by an unitary on all the previous collected outputs $\{S_{\ell}^\prime\}_{\ell=1}^k$ and an arbitrary number of ancilla in an arbitrary state.

The final decision of an adaptive protocol $\mathbb{P}_u$ is obtained from a measurement on the final outputs $\bm \rho=\{\rho_{\calE_n,u}\}_{n=1}^m$ (combining the prior $\bm P$)---the problem of channel discrimination is reduced to a state discrimination problem. Therefore, for a fixed protocol $\mathbb{P}_u$, one can introduce the same set of performance metrics: the inconclusive probability $P_F$ in Eq.~\eqref{PF_def} and the error probability constraints in Eq.~\eqref{epsilon_constraints} or Eq.~\eqref{epsilon_constraints_R}. However, the minimum inconclusive probability for $\bm \epsilon$-approximate UD is now a constrained minimization over all sensing protocols, 
\be 
P_{F,u}^{X\star}\left(\bm \epsilon; \{\bm \calE,\bm P\}\right)=\min_{\mathbb{P}_u:\mathbb{E}_X} P_F\left(\bm \epsilon; \{\bm \calE,\bm P\},\mathbb{P}_u\right),
\label{PF_u}
\ee 
for $X=U,R$ denoting the un-rescaled or rescaled constraints, which is even more challenging than the state case in Eq.~\eqref{PF_star_def}.
%With entanglement assistance, one can potentially boost the performance by considering an entangled pure state $\psi$ of the input and an ancilla, producing the states $\{\calE_n\otimes \calI (\psi)\}_{n=1}^m$ for the ensemble of channels. More generally, one can adopt an adaptive strategy~\cite{zhuang2020entanglement}.  
%In a general entanglement-assisted adaptive protocol $\mathbb{P}_u$ of $u$ channel uses, each operation can depend on all previous measurement results at earlier times. 
Similar to the state case, the minimum error probability of adaptive channel MED~\cite{pirandola2019fundamental,zhuang2020ultimate} can also be given by $
P_u(\{\bm \calE,\bm P\})=\min^\prime_{\bm \epsilon} \bm P\cdot \bm \epsilon,
$ 
where the minimization is under the constraint $P_{F,u}^{X\star}(\bm \epsilon;\{\bm \calE,\bm P\})=0$.

In Sec.~\ref{sec:LB_channel}, we give a general lower bound to $P_{F,u}^{X\star}$, valid for any $u$-round adaptive sensing protocol. Then we evaluate the lower bound for three examples in Sec.~\ref{sec:channel_example}. \QZ{We close this section by a discussion of the advantages from utilizing entanglement in the protocol.}

\subsection{Lower bound on the ultimate limit of inconclusive probability}
\label{sec:LB_channel}

\begin{figure}
\centering
\includegraphics[width=0.5\textwidth]{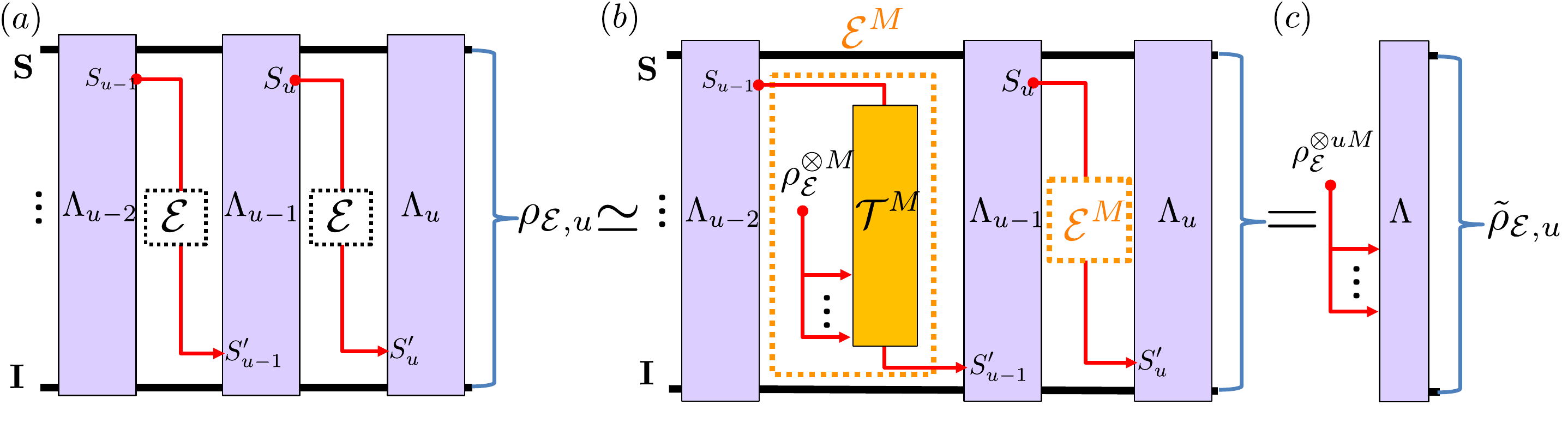}
\caption{Schematic of a $u$-step adaptive protocol $\mathbb{P}_u$ for channel discrimination.
\label{fig:schematic_general}
}
\end{figure}

To prove a lower bound of the inconclusive probability, we need to optimize over all possible protocols. To avoid such a challenging optimization, for each protocol $\mathbb{P}_u$, following Refs.~\cite{pirandola2019fundamental,zhuang2020ultimate} we can design another protocol to simulate its action. Suppose the protocol $\mathbb{P}_u$ on the $d$-dimensional channel $\calE$ produces state $\rho_{\calE,u}$, the simulation protocol produces an approximation $\tilde{\rho}_{\calE,u}$ of $\rho_{\calE,u}$, only relying on multiple copies of the Choi state
\be 
\rho_\calE=\left(\calE\otimes \calI\right) \zeta,
\ee
of the channel $\calE$ rather than utilizing the original channel. Here $\calI$ is the identity channel and the pure state 
$
\ket{\zeta}= \sum_{\ell=0}^{d-1} \ket{\ell,\ell}/\sqrt{d}
$ 
is a $d$-dimensional maximally-entangled state. The protocol (see Fig.~\ref{fig:schematic_general}(b)(c)) combines two techniques: channel simulation~\cite{quantumPQGA,PLOB,TQCtheory,banchi2020convex} and protocol stretching~\cite{PLOB,TQCtheory}.

First, we consider each channel use of $\calE$ in the original sensing protocol. The action of channel $\calE$ on the input can in general be approximated by a universal programmable quantum processor, given some program state that describes the channel $\calE$~\cite{banchi2020convex}. Note that here the quantum processor does not contain any information about the channel $\calE$. In this paper, we consider processor as the general teleportation operation $\calT^M$, as depicted in Fig.~\ref{fig:schematic_general}(b), while the program state is $M \ge 1$ copies of the Choi states $\rho_\calE$. We denote such an approximation of the input-output relation as a channel $\calE^M$. 

In general, the teleportation operation $\calT^M$ can be chosen as port-based teleportation (PBT)~\cite{ishizaka2008asymptotic} and the number of Choi states $M$ can be optimized to obtain the best bound. In the special case of teleportation-covariant channels, where for each Pauli unitary $U$ and any quantum state $\rho$, one can find another unitary $V$ such that
\be 
\calE\left(U \rho U^\dagger\right)=V \calE\left(\rho\right) V^\dagger,
\label{tele_covariant}
\ee 
one can simply use the direct teleportation to achieve an exact simulation with $m=1$ copy of Choi state.

The precision of such a channel simulation is quantified by the diamond norm deviation 
\be 
\Delta_{\calE,M}\equiv \|\calE-\calE^M\|_\diamond,
\ee 
where $\|A\|_\diamond=\sup_{\rho}\|A\otimes \calI \left(\rho\right)\|$ is the diamond norm~\cite{KitaevDiamond,PaulsenBook}. For teleportation-covariant channels, the simulation error $\Delta_{\calE,M}=0$ for $M=1$; while for PBT simulation, the simulation error can in general be bounded by~\cite{pirandola2019fundamental}
\begin{equation}
\Delta_{\calE,M} \le \delta_{M,d} \equiv 2d(d-1) M^{-1},\label{maxDiamond}
\end{equation}
which is valid for any number of ports $M \ge 1$ and any input dimension $d \ge 2$ for the channel. In general, other simulation protocols can also be used, and may potentially further decrease the simulation error and improve our overall lower bound.

By replacing the channel $\calE$ with $\calE^M$ in each step, one can obtain an overall protocol approximating the actual sensing protocol. Using the triangle inequality repeatedly, we can bound the deviation between the output state $\rho_{\calE,u}$ of the actual protocol and the output state $\tilde{\rho}_{\calE,u}$ of the simulated protocol as follows
\be
\|\rho_{\calE,u}-\tilde{\rho}_{\calE,u}\| \le u \Delta_{\calE,M}.\label{errorPROP}
\ee
To provide a lower bound on the sensing performance, the final step is protocol stretching~\cite{PLOB,pirandola2017ultimate,pirandola2019fundamental}. First, we notice that in each step of the protocol, the only channel-dependent component is the $M$ copies of the Choi states. Moreover, the overall $uM$ copies of the Choi states consumed in the $u$ steps can be directly prepared at the beginning of the entire protocol. Therefore, as depicted in Fig.~\ref{fig:schematic_general}(c), the simulated protocol can be organized into an equivalent protocol $\Lambda$ on the $uM$ copies of Choi states, so that the approximate output state $\tilde{\rho}_{\calE,u}$ is can be obtained as $\tilde{\rho}_{\calE,u}=\Lambda(\rho_\calE^{\otimes u M})$.
Combining this with Eq.~(\ref{errorPROP}) we have the overall error
\be
\|\rho_{\calE,u}-\Lambda(\rho_\calE^{\otimes u M})\| \le u \Delta_{\calE,M}.
\label{stretchingEQ}
\ee
With the above preparation, we can prove the following ultimate limit of approximate unambiguous channel discrimination.
\begin{theorem}
\label{theorem:LB}
Consider arbitrary $m \ge 2$ $d-$dimensional quantum channels $\bm \calE=\{\calE_n\}_{n=1}^{m}$ with prior probabilities $\bm P=\{p_n\}_{n=1}^{m}$. The failure probability of an arbitrary $u$-step adaptive protocol for $\bm \epsilon$-approximate UD satisfies
\begin{align}
&P_{F,u}^{U\star}\left(\bm \epsilon;\{\bm \calE,\bm P\}\right)\ge 
P_{F,u,LB,M}^U\left(\bm \epsilon;\{\bm \calE,\bm P\}\right)
\nonumber
\\
&\equiv P_F^{U\star}\left(\bm \epsilon +u\Delta_{\bm\calE,M}; \{\rho_{\bm\calE}^{\otimes u M},\bm P\}\right)-u\overline{\Delta}_M/2,
\label{bound2}
\end{align}
where we have used the notation $\Delta_{\bm\calE,M}\equiv\{u \Delta_{\calE_n,M}\}_{n=1}^m$ and $\rho_{\bm\calE}^{\otimes u M}\equiv \{\rho_{\calE_n}^{\otimes u M}\}_{n=1}^m$. Both the average simulation error $\overline{\Delta}_M=\sum_n P_n \Delta_{\calE_n,M}$ and each $\Delta_{\calE_n,M}$ in $\Delta_{\bm\calE,M}$ can be replaced by the uniform error $\delta_{M,d}$ of Eq. (\ref{maxDiamond}).
\end{theorem}

In general, one optimizes $P_{F,u,LB,M}$ in Eq.~\eqref{bound2} over $M$ to obtain the best lower bound. It is open whether this lower bound can be achievable in general. For the special case of a jointly-teleportation-covariant ensemble of channels, where each channel $\calE_n\in \calE$ satisfies the relation in Eq.~\eqref{tele_covariant} with the same choice of $V$ for each $U$, the simulation becomes exact at $M=1$, therefore we have the following corollary.
\begin{corollary}
\label{coro:tele}
Consider arbitrary $m \ge 2$ jointly-teleportation-covariant channels $\bm \calE$ with prior probabilities $\bm P$. The minimum failure probability of an arbitrary $u$-step adaptive protocol for $\bm \epsilon$-approximate UD equals the corresponding formula computed over their Choi matrices
\be
P_{F,u}^{U\star}\left(\bm \epsilon;\{\bm \calE,\bm P\}\right)=P_F^{U\star}(\bm \epsilon; \{\rho_{\bm\calE}^{\otimes u },\bm P\}).
\label{bound2tele}
\ee
This is achievable by a non-adaptive entanglement-based strategy
where $u$ copies of a maximally-entangled state $\zeta $ are sent through the
extended channel $\mathcal{E}_{n}\otimes \mathcal{I}$.
\end{corollary}
This means that for jointly-teleportation-covariant channels, adaptivity is not necessary. As a by-product of our lower bound, we can extend the results of Ref.~\cite{wang2006unambiguous} to include adaptive strategies for teleportation-covariant channels---the $\bm \epsilon=0$ case of Eq.~\eqref{bound2tele} corresponds to the exact UD case. In fact, because for teleportation-covariant channels the simulations are exact, therefore Corollary~\ref{coro:tele} can be extended to the rescaled case
\be
P_{F,u}^{R\star}\left(\bm \epsilon;\{\bm \calE,\bm P\}\right)=P_F^{R\star}(\bm \epsilon; \{\rho_{\bm\calE}^{\otimes u },\bm P\}).
\label{bound2tele_R}
\ee

Although Eq.~\eqref{bound2} in general reduces the channel problem to discrimination between Choi states, solving the minimum inconclusive probability of approximate UD for general mixed states still requires challenging optimizations. Here we make use of the lower bound from Eq.~\eqref{PF_lower_bonund_exact} of Lemma~\ref{lemma:fidelity_LB} and Eq.~\eqref{PF_connections}, and further derive the following lower bound for binary mixed states.
\begin{lemma}
\label{lemma:bound_fidelity_channel}
Consider a pair of nonidentical channels $\{\calE_p,\calE_q\}$ with prior $\{p,q\}$, the minimum failure probability as a function of the un-rescaled tolerance $\{\epsilon_p^U,\epsilon_q^U\}$ satisfies
\begin{align}
&P_{F,u}^{U\star}\ge 
P_{F,u,LB,M}^U
\equiv 
g(F^{uM}_{\calE_p,\calE_q};\epsilon_p^R,\epsilon_q^R)-u\overline{\Delta}_M/2,
\label{bound_fidelity_channel}
\end{align}
with $\left\{\epsilon_p^R,\epsilon_q^R\right\}$ obtained from solving $\{\epsilon_p^U+u \Delta_{\calE_p,M},\epsilon_q^U+u \Delta_{\calE_q,M}\}=\left[1-g\left(F^{uM}_{\calE_p,\calE_q};\epsilon_p^R,\epsilon_q^R\right)\right]\left\{\epsilon_p^R,\epsilon_q^R\right\}$.
\end{lemma}
Here we utilized $F\left(\rho_{\calE_p}^{\otimes u M},\rho_{\calE_q}^{\otimes u M}\right)=F^{uM}\left(\rho_{\calE_p},\rho_{\calE_q}\right)\equiv F^{uM}_{\calE_p,\calE_q}$, where we introduced $F_{\calE_p,\calE_q}$ as the fidelity between the Choi matrices for the two channels. The function $g$ is defined by Eq.~\eqref{PF_opt_main_g}. Note that when $p=q=1/2$, we can switch function $g$ in Eq.~\eqref{bound_fidelity_channel} to the analytical function $h$ given by Eq.~\eqref{PF_lower_bonund_h} to have an analytical bound.

%\QZE{**This might be wrong**}
Similar to Corollary \ref{coro:tele}, when the two channels are teleportation-covariant, we can let $M=1$, $\Delta_{\calE_p,M}=\Delta_{\calE_q,M}=0$ and obtain a simplified bound
\begin{align}
&P_{F,u}^\star\ge 
P_{F,u,LB}^U
=g(F^{u}_{\calE_p,\calE_q};\epsilon_p^R,\epsilon_p^R)
\label{bound_fidelity_channel_tele}
\end{align}
with $\left\{\epsilon_p^R,\epsilon_q^R\right\}$ obtained from solving $\{\epsilon_p^U,\epsilon_q^U\}=\left[1-g\left(F^{u}_{\calE_p,\calE_q};\epsilon_p^R,\epsilon_q^R\right)\right]\left\{\epsilon_p^R,\epsilon_q^R\right\}$.

\subsection{Examples}
\label{sec:channel_example}

\begin{figure}
\centering
\includegraphics[width=0.45\textwidth]{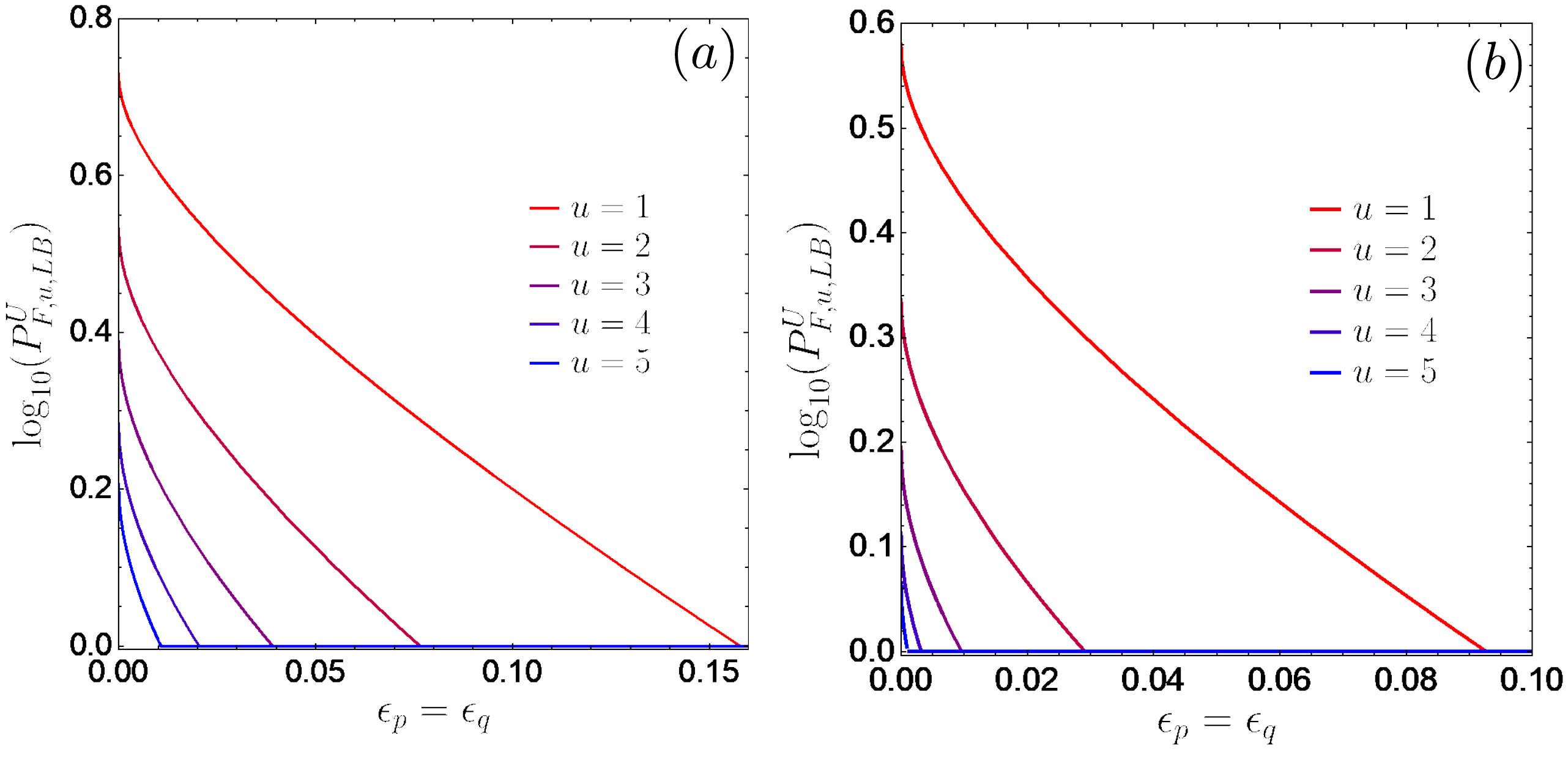}
\caption{
Lower bound $P^U_{F,u,LB}$ vs the error tolerance for approximate UD between quantum channels of symmetric prior $p=q=1/2$. (a) Noisy Pauli gates with $\eta=0.6$,  (b) Quantum erasure channels with $\braket{e_1|e_2}=0.3$.
\label{fig:LB_TC}
}
\end{figure}

Now we evaluate the lower bound Eq.~\eqref{bound_fidelity_channel} in Lemma~\ref{lemma:bound_fidelity_channel} in three examples. We first consider jointly-teleportation-covariant channels, including noisy Pauli gates and quantum erasure channels. For these cases, Lemma~\ref{coro:tele} can be directly utilized and upper bounds can also be obtained by designing measurement schemes on the Choi states. Then we consider the general case of amplitude-damping channels, which are not teleportation-covariant.

{\em Noisy Pauli gates.---}
Single-qubit gates are a fundamental building block in quantum computers. It is therefore important to certify the type of gates being built. Practical gates are noisy, which makes the discrimination between different gates challenging. As an example, we consider the discrimination between  two single-qubit Pauli gates in presence of depolarizing noise
\be 
\calE_k(\rho)=\eta \sigma_k \rho \sigma_k+\left(1-\eta\right)\frac{I_2}{2}, k=Z,I,
\label{channel_Pauli}
\ee 
with $\sigma_Z$ being the Pauli operator and $\sigma_I=I_2$ being identity. As Pauli channels, $\calE_Z$ and $\calE_I$ are jointly-teleportation-covariant, with Choi states
\be 
\rho_{\calE_k}=\eta \state{\Phi_k}+(1-\eta)\frac{I_4}{4},
\label{Choi_Pauli}
\ee 
where $\ket{\Phi_I}=\ket{+}$ and $\ket{\Phi_Z}=\ket{-}$ are maximally entangled. We see that the Choi states in Eq.~\eqref{Choi_Pauli} are identical to the binary mixed states with depolarizing noise described by Eq.~\eqref{rho_pm}.
From Corollary~\ref{coro:tele}, the results in Sec.~\ref{sec:data_processing} immediately provide upper and lower bounds for the $u=1$ case of channel discrimination, as shown in Fig.~\ref{fig:PF_epsilon}(a). From the fidelity in Eq.~\eqref{fideility_example_1}, we can also obtain the lower bound in Eq.~\eqref{bound_fidelity_channel_tele} for different $u$, as shown in Fig.~\ref{fig:LB_TC}(a).

{\em Quantum erasure channels.---} 
We consider two different quantum erasure channels. Upon input $\rho$, each channel $\calE_k$ gives the output
\be
\calE_{k}(\rho)=\eta \state{e_k}+(1-\eta)\rho,
\ee 
on the input $\rho$ for $k=1,2$, with two pure error states $\{\ket{e_1},\ket{e_2}\}$ orthogonal to the input Hilbert space. Note that $\braket{e_1|e_2}$ can be non-zero in general. The two channels are jointly-teleportation-covariant, as the teleportation unitaries act only on the input Hilbert space. The Choi states
\be 
\rho_{\calE_k}=\eta \state{e_k}_E\QZ{\otimes \frac{I_2}{2}}+(1-\eta)\state{\zeta}_{SI}
\label{channel_QEC}
\ee 
differ only by the error part, similar to the binary mixed state with a single common component considered in Eq.~\eqref{states_example_2}. Therefore, from Corollary~\ref{coro:tele}, the upper bounds and lower bounds in Sec.~\ref{sec:data_processing} can be utilized for the $u=1$ case here, which are calculated in Fig.~\ref{fig:PF_epsilon}(b). From the fidelity in Eq.~\eqref{fideility_example_2}, we can also obtain the lower bound in Eq.~\eqref{bound_fidelity_channel_tele} for different $u$, as shown in Fig.~\ref{fig:LB_TC}(b).

\begin{figure}
\centering
\includegraphics[width=0.5\textwidth]{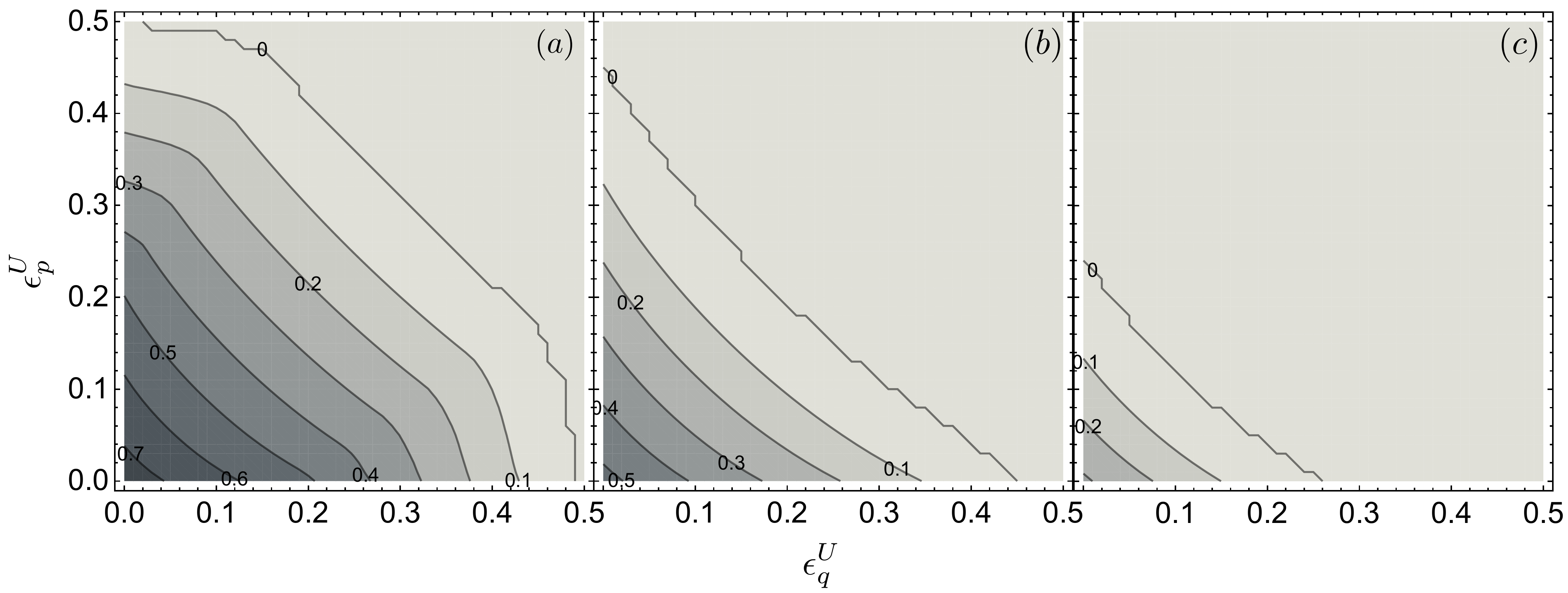}
\caption{Optimum lower bound $\max_M P_{F,u,LB,M}^U$ (Eq.~\eqref{bound_fidelity_channel} of Lemma~\ref{lemma:bound_fidelity_channel}) for a pair of quantum amplitude damping channels with damping parameters $r_q=0.8,r_p=0.9$.
(a)$u=1$, (b)$u=2$, (c)$u=3$.
We numerically scan over $\epsilon_p^R,\epsilon_q^R$ to obtain both $\epsilon_p^U,\epsilon_q^U$ and $P_{F,u,LB,M}^U$; then we optimize over $M$ for each $\epsilon_p^U,\epsilon_q^U$. The optimum $M$ is given in the Appendix,
\label{fig:dephasing_LB}
}
\end{figure}

\begin{figure}
\centering
\includegraphics[width=0.4\textwidth]{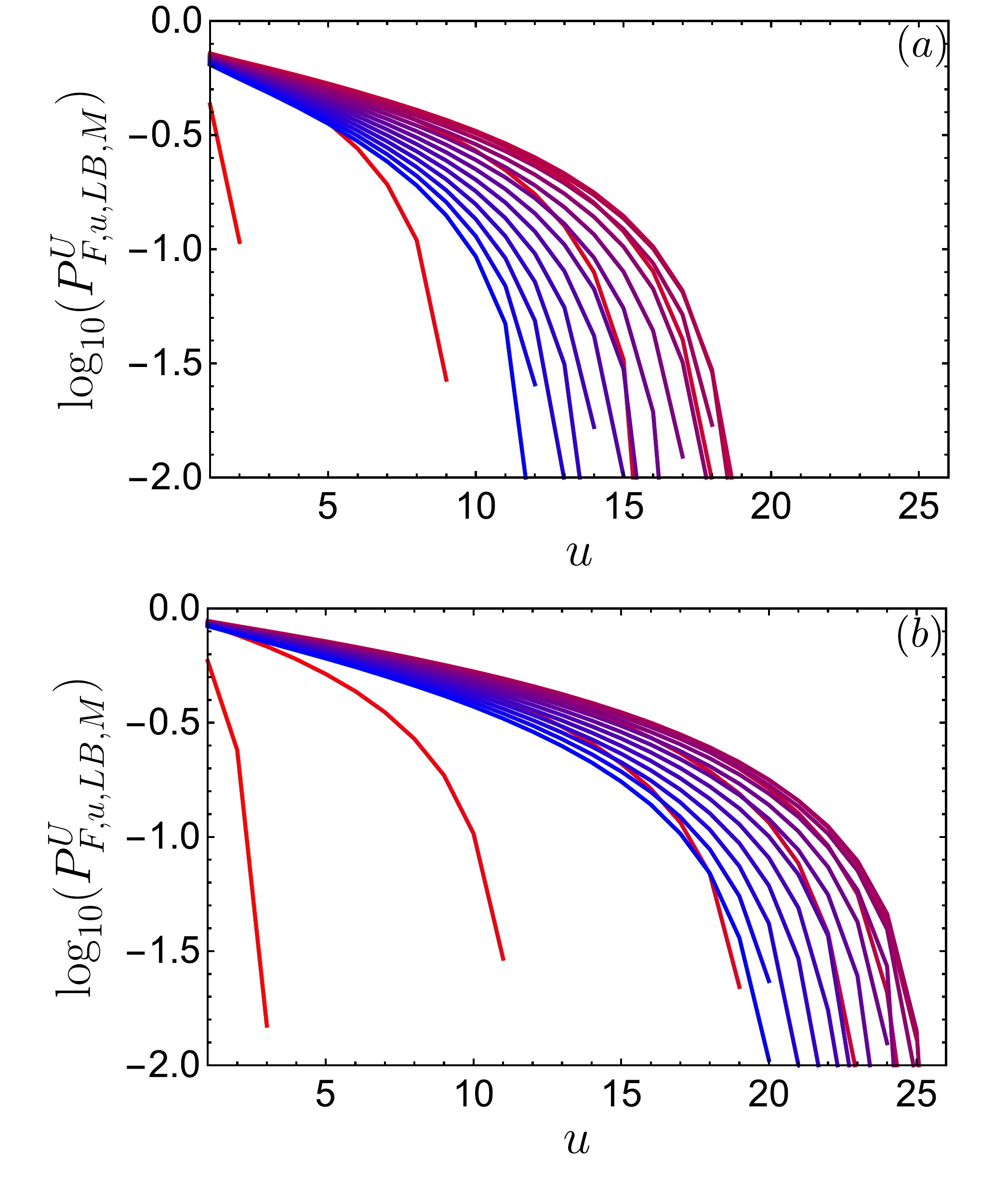}
\caption{Lower bound $ P_{F,u,LB,M}^U$ in Eq.~\eqref{bound_fidelity_channel} of lemma~\ref{lemma:bound_fidelity_channel} vs. the number of rounds $u$. 
$r_q=0.98,r_p=0.99$.
(a) $\epsilon_q^U=0.1,\epsilon_p^U=0.1$
(b) $\epsilon_q^U=0.02,\epsilon_p^U=0.04$.
Color indicates $M$, from red ($M=1$) to blue ($M=81$), in steps of $\Delta M=5$.
\label{fig:dephasing_LB_change_u}
}
\end{figure}

{\em Amplitude-damping channels.---} A quantum amplitude damping channel $\mathcal{A}_{r}$\ with damping probability $r$\ has the Kraus representation $\mathcal{A}%
_{r}(\rho )=\sum_{i=0,1}K_{i}\rho K_{i}^{\dagger }$, with operators $K_{0}=\ketbra{0}{0}+\sqrt{1-r}\ketbra{1}{1}$ and $K_{1}=\sqrt{r}\ketbra{0}{1}$. It
is not tele-covariant and its PBT simulation has non-zero error $\Delta _{\mathcal{A}_{r},M}=\xi_{M}[\left( 1-r\right) /2+\sqrt{1-r}],$ where $\xi
_{M}$ is the constant given in Ref.~\cite[Eq.~(11)]{pirandola2019fundamental}. We consider the $u$-round approximate UD between $\mathcal{A}_{r_q}$ and $\mathcal{A}_{r_p}$. The fidelity between the Choi states of the channels can be obtained analytically~\cite{pirandola2019fundamental,zhuang2020ultimate}
\be 
F_{\mathcal{A}%
_{r_q},\mathcal{A}%
_{r_p}}\equiv\left[ 1+\sqrt{(1-r_q)(1-r_p)}+\sqrt{r_qr_p}\right] /2.
\ee 
We can therefore evaluate the lower bound $P_{F,u,LB,M}^U$ in Eq.~\eqref{bound_fidelity_channel} of Lemma~\ref{lemma:bound_fidelity_channel}. First, we optimize over $M$ and calculate the optimum lower bound $\max_M P_{F,u,LB,M}^U$ for different error constraints and number of rounds $u$ in Fig.~\ref{fig:dephasing_LB}. We see that as the number of rounds $u$ increases, the lower bound is decreasing as expected. To further understand the trend, we also evaluate the change of the lower bound $P_{F,u,LB,M}^U$ with $u$ for different fixed $M$ in Fig.~\ref{fig:dephasing_LB_change_u}. The envelop of the lower bounds for all $M$ is the optimum lower bound we considered in Fig.~\ref{fig:dephasing_LB}. We see that when $M$ is small (red lines), the simulation error is large and causes the lower bound to be small; while when $M$ is large (blue lines), the abundance of Choi states makes the lower bound again small. For each $u$, one has an optimum $M$ to balance between the above two effects and provide the tightest lower bound.

\QZ{
{\em Entanglement advantages over classical schemes.---}To better understand the effects of entanglement, here we consider non-adaptive classical strategies without entangled ancilla. To show the advantage of entanglement, we need upper bounds on the inconclusive probability of the entangled strategy and lower bounds on the classical correspondence. While efficient calculable lower bounds are provided in Lemma~\ref{lemma:bound_fidelity_channel}, upper bounds on the performance are hard to obtain, even for the approximate UD between the Choi states. Only in the $u=1$ case, we have upper bounds available, as depicted in Fig.~\ref{fig:PF_epsilon}. Therefore, we focus on the $u=1$ case.
}

\QZ{
First, we consider the noisy Pauli gates specified in Eq.~\eqref{channel_Pauli}. From the symmetry of the channel, we consider the input $\ket{+}$ in the classical strategy. Then one reduces the problem to states
\be 
\rho_{Z}=\eta \state{-}+\frac{(1-\eta)}{2} I_2,
\rho_{I}=\eta \state{+}+\frac{(1-\eta)}{2} I_2.
\ee  
The fidelity between the states
$
F\left(\rho_{Z},\rho_{I}\right)=\sqrt{1-\eta^2}.
$ 
From Ineq.~\eqref{bound_fidelity_channel_tele}, the lower bound of the inconclusive probability can be obtained 
\begin{align}
&P_{F,u}^\star\ge 
P_{F,u,LB}^U
=g(\sqrt{1-\eta^2};\epsilon_p^R,\epsilon_p^R).
\label{LB_classical}
\end{align}
We compare the above classical lower bound with the upper bound of the entangled strategies in Fig.~\ref{fig:PF_epsilon}. In the range around $0.07\lesssim \epsilon_p^U=\epsilon_q^U\lesssim 0.2$, we can see an advantage from entanglement. We expect better entangled strategies to exist within the shaded region and a larger advantage is possible. The entanglement's benefit in presence of depolarizing noise resembles the advantage in quantum illumination~\cite{tan2008quantum}. Despite entanglement being fragile, its advantage over the performance achievable with only initial classical correlations survives noise.
}

\QZ{
Next we consider the erasure channels specified in Eq.~\eqref{channel_QEC}. From the symmetry of the channel, we can simply consider the input $\ket{0}$, leading to the output 
\be 
\rho_{k}=\eta \state{e_k}_E+(1-\eta)\state{0},
k=1,2.
\ee 
The fidelity can be calculated as 
$
F=\eta |\braket{e_1|e_2}|+(1-\eta),
$ 
which is identical to the case with entanglement assistance. Therefore, the lower bound of the classical strategy coincides with the ultimate lower bound. And we are not able to show any entanglement advantage in this case, similar to the MED case in Ref.~\cite{zhuang2020ultimate}.
}

\section{Conclusions}

In this paper, we formulated the approximate unambiguous discrimination scenario for quantum states and quantum channels. For the binary pure states case, we are able to solve the minimum inconclusive probability as a function of the conclusive error probability constraints for an arbitrary prior. The minimum inconclusive probability satisfies convexity, continuity and data-processing, which makes it more friendly for both theory analyses and experimental realizations. For the channel case, we are able to prove an ultimate lower bound of the minimum inconclusive probability for any adaptive sensing protocols, based on the approximate unambiguous discrimination between the Choi states. For jointly-teleportation-covariant channels, the lower bound can be achieved with maximum entangled inputs and no adaptive strategy is required.

\begin{acknowledgements}
Q.Z. acknowledges funding from Defense Advanced Research Projects Agency (DARPA) under Young Faculty Award (YFA) Grant No.
N660012014029 and University of Arizona. Q.Z. thanks Stefano Pirandola and Jarom\'{i}r Fiur\'{a}\v{s}ek for discussions.
\end{acknowledgements}

\appendix 
\section{Binary pure states}
\label{sec:binary_pure_state_details}

%From symmetry we must have the solution
%\be 
%P_F^{R\star}\left(\{\epsilon_p,\epsilon_q\};\{\{\ket{p},\ket{q}\},\{p,q\}\}\right)=g(|\braket{p|q}|;\epsilon_p,\epsilon_q,p,q)
%\ee 
%also it should increase as $|\braket{p|q}|$ increases.

In the first example of binary pure states discrimination, we are able to solve $P_F^{R\star}(\bm \epsilon;\bm \Upsilon)$ as a function of $\bm \epsilon$ for an arbitrary prior. The equal prior exact UD is sovled in Refs.~\cite{dieks1988overlap,jaeger1995optimal}. Here, we consider a general prior $\bm P=\{p,q\}$ and solve the approximate UD. Following Ref.~\cite{dieks1988overlap}, we adopt a measurement protocol assisted by an ancilla qubit in a pure state $\ket{s_0}$. A general unitary is applied on the ancilla and input to transform
\begin{align}
&\ket{p s_0}\to \alpha \ket{p_1 s_1}+\beta \ket{p_2 s_2},
\\
&\ket{q s_0}\to \gamma \ket{q_1 s_1}+\delta \ket{q_2 s_2},
\end{align}
such that the output states of the ancillar is orthogonal, i.e., $\braket{s_1|s_2}=0$. Then by measuring the ancilla in the $\{\ket{s_1},\ket{s_2}\}$ bases, we can determine inconclusive when the outcome is $s_2$ and continue to perform a second measurement on the input when the outcome is $s_1$.

While Ref.~\cite{dieks1988overlap} considers $\braket{p_1|q_1}=0$ to enable exact UD, we let $\braket{p_1|q_1}\ge0$ in general. We consider a projective measurement along $\ket{\phi_\theta}=\cos\theta \ket{p_1}+\sin\theta \ket{p_1^\perp}$ and $\ket{\phi_\theta^\perp}=\sin\theta \ket{p_1}-\cos\theta \ket{p_1^\perp}$ to trade-off the error probabilities such that
\begin{align}
&P_{E|p}=|\braket{p_1|\phi_\theta^\perp}|^2=\sin^2\theta \le \epsilon_p
\\
&P_{E|q}=|\braket{q_1|\phi_\theta}|^2
=|\cos\theta \braket{q_1|p_1}+\sin\theta \braket{q_1|p_1^\perp}|^2\le \epsilon_q.
\end{align}
We can parameterize $\braket{q_1|p_1^\perp}=\sqrt{1-|\braket{q_1|p_1}|^2}e^{i \alpha}$. The first constraint is that there is a valid solution for $\theta,\alpha$, which leads to the constraint
\be
|\braket{p_1|q_1}|\in [\epsilon_{pq-},\epsilon_{pq+}],
\ee
where
\be 
\epsilon_{pq\pm}=
|\sqrt{\epsilon_p(1-\epsilon_q)}\pm \sqrt{\epsilon_q(1-\epsilon_p)}|.
\ee

The other constraint is unitarity, which preserves the inner product,
\be 
\braket{p|q}=\alpha^\star\gamma \braket{p_1|q_1}+\delta^\star \beta \braket{q_2|p_2}.
\label{inner_prod}
\ee  
Because of norm preservation, $|\alpha|^2+|\beta|^2=|\gamma|^2+|\delta|^2=1$.
Under these constraints, we want to minimize
\be 
P_F=p|\beta|^2+q |\delta|^2.
\ee 

\begin{figure}
\centering
\includegraphics[width=0.3\textwidth]{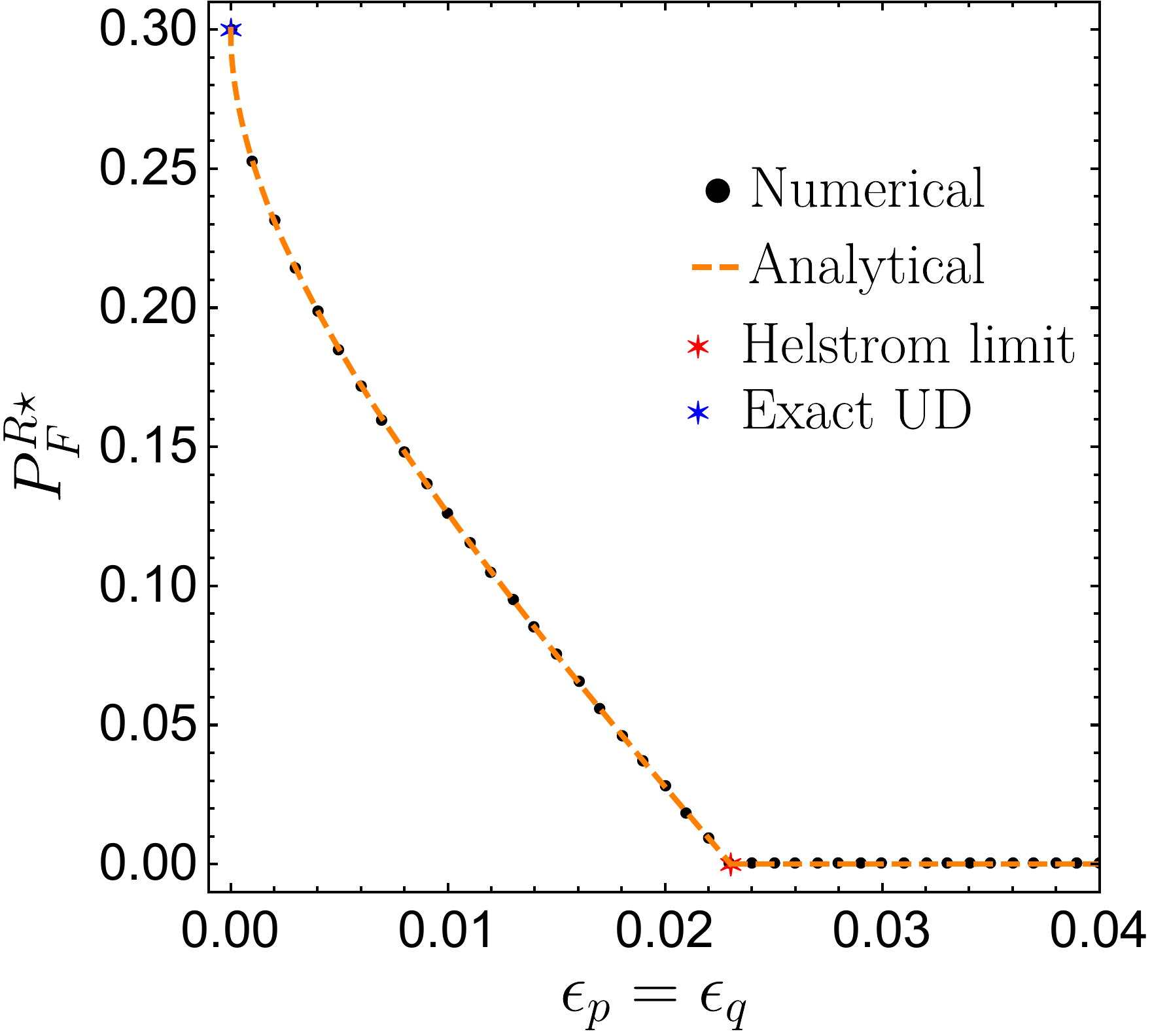}
\caption{$P_F^{R\star}$. Comparison between numerical results and analytical results in the symmetric case of $p=q=1/2$ and $\braket{p|q}=0.3$. 
\label{fig:PF_epsilon_confirm}
}
\end{figure}

It is straightforward to see that to minimize $P_F$, we need $|\braket{q_2|p_2}|=1$ and the constraint in Eq.~\eqref{inner_prod} becomes
\begin{align}
&|\braket{p|q}|=\sqrt{1-|\beta|^2}\sqrt{1-|\delta|^2} |\braket{p_1|q_1}|+|\delta| |\beta|. 
\end{align}

We can re-parameterize with angle parameters in $[0,\pi/2]$ as $\beta=\sin \tilde{\beta}, \delta=\sin \tilde{\delta}$. Then the we can obtain the optimization problem in Eq.~\eqref{PF_opt_main_g}.
We can use inequality, 
\be 
p \sin^2 \tilde{\beta}+q \sin^2 \tilde{\delta}\ge 2\sqrt{pq} \sin \tilde{\beta}\sin\tilde{\delta},
\label{min1}
\ee 
which is achievable when $p \sin^2 \tilde{\beta}=q \sin^2 \tilde{\delta}$. We can obtain a lower bound that leads to the solution
in Eq~\eqref{PF_lower_bonund_h} via Lagrangian multiplier methods. Moreover, the only case that both Eq.~\eqref{min1} and~\eqref{PF_lower_bonund_h} can be achieved is $p=q=1/2$, where we have
\be 
P_F^{R\star}=1-\frac{1-|\braket{p|q}|}{1-\epsilon_{pq+}}.
\label{PF_Sym}
\ee 
In fact, one can check that this result agrees with Ref.~\cite{chefles1998strategies}.

We can check both end points analytically. When $P_F=0$, we have $\tilde{\alpha}=\tilde{\beta}=0$ and $|\braket{p_1|q_1}|=\braket{p|q}$, then simple optimization gives the Helstrom limit; In particular, for the equal-prior case, Eq.~\eqref{Helstrom} leads to $P_H=\min (\epsilon_p^R+\epsilon_q^R)/2$ under the constraint $|\braket{p|q}|=\epsilon_{pq+}$. From symmetry, it is achieved at $P_H=\epsilon_p=\epsilon_q=\left(1-\sqrt{1-4|\braket{p|q}|^2}\right)/2$ which agrees with the Helstrom limit. When $\epsilon_p^R=\epsilon_q^R=0$, we can show that $\braket{q_1|p_1}=0$ and then the problem goes back to the exact UD case; and for the symmetric case one can obtain the exact UD results in Ref.~\cite{peres1988differentiate}.

To validate our numerical results of Eqs.~\eqref{PF_opt_main_g}, in Fig.~\ref{fig:PF_epsilon_confirm} we consider the equal prior case and compare the numerical results (black dots) and the analytical results of Eq.~\eqref{PF_Sym} (orange dashed) along the line of $\epsilon_p=\epsilon_q$. A perfect agreement is found. Moreover, we analytically calculate the end points of exact UD (blue circle) and Hesltrom limit (red circle) and they also agree well with our results.

\section{A summary of previous results}

In Ref.~\cite{feng2004unambiguous}, it was proven that $\bm \rho$ can be exact unambiguously discriminated if and only if for any $1\le n\le m$, ${\rm supp}(\bm \rho)\neq {\rm supp}(\bm \rho\setminus \rho_n)$. Here ${\rm supp}(\rho)$ is the support of the state $\rho$. For pure states, this reduces to the previously known condition of linear independence~\cite{chefles1998unambiguous}. For multiple copies, Ref.~\cite{wang2006unambiguous} shows that: if for any $n\neq n^\prime$, ${\rm supp}(\rho_n)\subsetneq {\rm supp}(\rho_{n^\prime})$, then $\{\rho_n^{\otimes m}\}_{n=1}^m$ can be unambiguously discriminated; Otherwise, for arbitrary $N$, $\{\rho_n^{\otimes N}\}_{n=1}^m$ cannot be unambiguously discriminated. Ref.~\cite{chefles1998optimum} solved the optimum exact UD cyclic symmetric pure states, with explicit measurement called equal-probability measurement (EPM). In Ref.~\cite{raynal2003reduction} the exact case between binary mixed states is reduced to minimum error discrimination. A neat geometric picture of UD is provided by Ref.~\cite{bergou2012optimal}. Many other special cases have been considered~\cite{herzog2007optimum,pang2009optimum,kleinmann2010structural,bergou2012optimal,bandyopadhyay2014unambiguous}.

Much less is known about the channel case.  In the UD case, previous results only consider the non-adaptive case. In Ref.~\cite{wang2006unambiguous}, sufficient and necessary condition for UD using non-adaptive strategies is given. By generalizing the support of states to channels as ${\rm supp}\left(\calE\right)=\{\sum_k \lambda_k E_K, \lambda_k\in \mathbb{C}\}$ (where $E_k$'s are the Kraus operators of the channel $\calE$), the single-use condition of exact UD is: for any $1\le n \le m$, ${\rm supp}(\calE_n)\subsetneq {\rm supp}(\bm\calE \setminus\calE_n)$. In non-adaptive multiple channel uses, it has been shown that: if for any $n\neq n^\prime$, ${\rm supp}( \calE_n)\subsetneq {\rm supp}(\calE_{n^\prime})$, then $\bm\calE$ can be unambiguously discriminated with $m$ uses; Otherwise, for arbitrary $N$ uses, $\bm\calE$ cannot be unambiguously discriminated.

For quantum states, when UD is not possible, various relaxations have been considered, as interpolation between ME and UD.
One can maximize the correct probability when conclusive, with a fixed probability of failure~\cite{chefles1998strategies,zhang1999general,eldar2003mixed,fiuravsek2003optimal,nakahira2015generalized,touzel2007optimal,nakahira2016finding}; one can consider the maximum-confidence criteria~\cite{croke2006maximum,herzog2009discrimination}, while trying to keep the failure probability as low as possible. Here confidence is the probability of the state being $n$ when the decision $n$ is made (note that it is different from the conditional error probability $\bm \epsilon$ considered in our paper); One can also generalize this approach to an error margin that can be continuously tuned~\cite{hayashi2008state}. One can also assign a cost function depending on the inconclusive probability and conclusive error probability~\cite{nakahira2015finding,combes2015cost,nakahira2016finding}, and then use algorithms to minimize the cost, which leads to solutions to general problems; Extensions to parameter-estimation scenarios have led to quantum metrology protocols assisted by abstention~\cite{fiuravsek2006optimal,gendra2013,gendra2013optimal}.

\section{Proof of Eq.~(\ref{PF_connections})} 
\label{PF_connections_proof}

Let $P_F^{U\star}(\bm \epsilon^U;\bm \Upsilon)=p_f$. Consider the re-scaled constraints
\be 
\bm \epsilon^R=\frac{1}{1-p_f}\bm \epsilon^U.
\ee
First, because the same choice of measurement can achieve $p_f$ while satisfying the rescaled constraint of $\bm \epsilon^R$, we have
\be 
P_F^{R\star}\left(\bm \epsilon^R;\bm \Upsilon \right)\le p_f.
\ee 
Next, if 
\be 
P_F^{R\star}\left(\bm \epsilon^R;\bm \Upsilon \right)=p_f^\prime<p_f, 
\label{proof_connection_step}
\ee 
then there is a corresponding
\be 
\bm \epsilon^{U\prime }=\left(1-p_f^\prime\right)\bm \epsilon^R=\frac{1-p_f^\prime}{1-p_f}\bm \epsilon^U>\bm \epsilon^U
\ee 
that guarantees $P_F^{U\star}(\bm \epsilon^{U\prime };\bm \Upsilon)\le p_f^\prime$. Here the inequality between vectors means a set of item-wise inequalities.
\begin{figure}[H]
\centering
\includegraphics[width=0.2\textwidth]{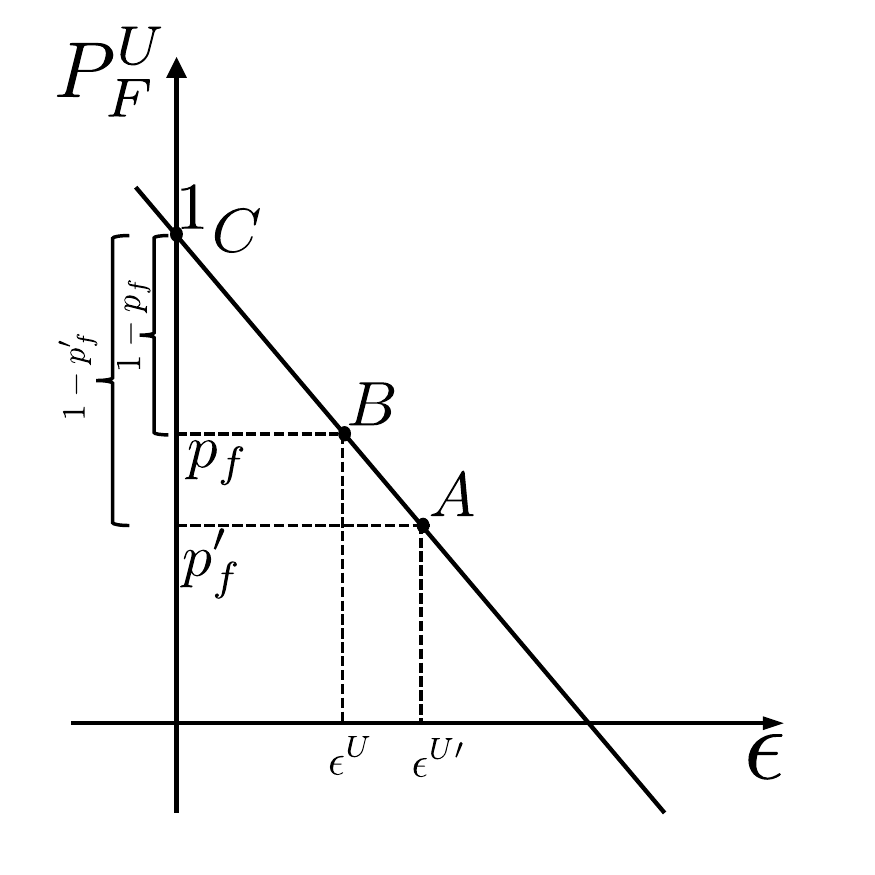}
\caption{Schematic of the proof of equivalence.
\label{fig:schematic_proof}
}
\end{figure}
Fig.~\ref{fig:schematic_proof} visualizes the setup by condensing all constraints into a single axis. The points $A=(\bm \epsilon^{U\prime },p_f^\prime)$, $B=(\bm \epsilon^{U },p_f)$ and $C=(\bm 0, 1)$ are on the same line. While $B$ point is on the curve $P_F^{U\star}$, the other two points $C$ and $A$ are upper bounds on $P_F^{U\star}$.
From convexity in Lemma~\ref{lemma:convexity} of the main paper, $C$ and $A$ have to be on the curve as well; namely, 
\be 
P_F^{U\star}(\bm \epsilon^{U\prime };\bm \Upsilon)= p_f^\prime, P_F^{U\star}(\bm 0;\bm \Upsilon)=1,
\ee 
and the straight line AC must be the optimum. If the $P_F^{U\star}(\bm \epsilon;\bm \Upsilon)$ is strictly convex in $\bm \epsilon$ or $P_F^{U\star}(\bm 0;\bm \Upsilon)<1$, then this leads to a contradiction, therefore the Ineq.~\eqref{proof_connection_step} cannot be true and 
\be 
P_F^{R\star}\left(\bm \epsilon^R;\bm \Upsilon \right)=p_f^\prime=p_f=P_F^{U\star}(\bm \epsilon^U;\bm \Upsilon).
\ee 
The only possible case would be a straight line $CA$ passing through $(0,1)$.

\section{Proof of Lemma~\ref{lemma:convexity} }
\label{app:convex_proof}

Suppose for each $\bm \epsilon^{(k)}$ the POVM achieving $P_F^{U\star}(\bm \epsilon^{(k)};\bm \Upsilon)$ is $\bm \Pi^{(k)\star}$.
Then we have
\be 
P_F^{U\star}(\bm \epsilon^{(k)};\bm \Upsilon)=\sum P_n \tr\left(\Pi_0^{(k)\star}\rho_n\right).
\ee 
The constraints give
\be 
P_{E|n}=1-\tr\left(\Pi_0^{(k)\star}\rho_n\right)-\tr\left(\Pi_n^{(k)\star}\rho_n\right)\le \epsilon^{(k)}_n.
\ee 
We can define POVMs $\bar{\bm \Pi}=\sum_k r_k\bm \Pi^{(k)\star}$, which represents a strategy that performs the measurement represented by $\bm \Pi^{(k)\star}$ with probability $r_k$. It is easy to verify $\sum_n \bar{\Pi}_n=\sum_k r_k I=I$ and they are positive semi-definite. Then the error probability
\begin{align}
P_{E|n}&=1-\tr\left(\sum_k r_k \Pi_0^{(k)\star}\rho_n\right)-\tr\left(\sum_k r_k\Pi_n^{(k)\star}\rho_n\right)
\nonumber
\\
&\le \sum_k r_k \epsilon^{(k)}_n.
\end{align} 
The inconclusive probability
\be
P_F=\sum P_n \tr\left(\sum_k r_k \Pi_0^{(k)\star}\rho_n\right)=\sum_k r_k P_F^{U\star}(\bm \epsilon^{(k)};\bm \Upsilon). 
\ee 
This strategy achieves the inconclusive probability $\sum_{k=1}^K r_k P_F^{U\star}(\bm \epsilon^{(k)};\bm \Upsilon)$ with error constraints $\sum_{k=1}^K r_k \bm \epsilon^{(k)}$, because better measurement can exist for the same error constraints, so we have Ineq.~\eqref{PF_ineq} of Lemma~\ref{lemma:convexity} in the main paper.

\section{Proof of Lemma~\ref{lemma:fidelity_LB}} 
\label{app:proof_fidelity_LB}
From data processing we have
\begin{align}
&P_F^{R\star}\left(\{\epsilon_p,\epsilon_q\};\{\{\rho_p,\rho_q\},\{p,q\}\}\right)\ge
\nonumber
\\
&P_F^{R\star}\left(\{\epsilon_p,\epsilon_q\};\{\{\psi_{\rho_p},\psi_{\rho_q}\},\{p,q\}\}\right)
\\
&=g(|\braket{\psi_{\rho_p}|\psi_{\rho_q}}|;\epsilon_p,\epsilon_q),
\end{align}
where we used the binary pure states result in Eq.~\eqref{PF_opt_main_g} of the main paper.
Similar to Ref.~\cite{rudolph2003unambiguous}, using Uhlmann's theorem that the maximum overlap between purification equals the fidelity $|\braket{\psi_{\rho_p}|\psi_{\rho_q}}|\le F\left(\rho_p,\rho_q\right)\equiv \tr \left(\sqrt{\sqrt{\rho_q} \rho_p \sqrt{\rho_q}}\right)$, also since $g$ is non-decreasing with $|\braket{\psi_{\rho_p}|\psi_{\rho_q}}|$ increasing, therefore one can obtain the tightest lower bound by taking 
\begin{align}
&P_F^{U\star}\left(\{\epsilon_p,\epsilon_q\};\{\{\rho_p,\rho_q\},\{p,q\}\}\right)\ge
\nonumber
\\
&P_{F,LB,1}^{R\star}=g(F\left(\rho_p,\rho_q\right);\epsilon_p,\epsilon_q).
\end{align}
Then further using the lower bound Eq.~\eqref{PF_lower_bonund_h} of the main paper for the binary pure states case, we can have the further lower bound $P_{F,LB,2}^{R\star}$.

\section{Proof of Lemma~\ref{approximate_continuity}}
\label{app:proof_approximate_continuity}
Here we extend the lemma to both rescaled and un-rescaled case and then prove both results.
\begin{lemma}
\label{approximate_continuity_full}
{\it Continuity of approximate UD:} Consider two set of states $\bm \rho=\{\rho_n\}_{n=1}^m$ and $\bm \rho^\prime=\{\rho_n^\prime\}_{n=1}^m$, with relative deviation
$
\|\rho_n-\rho_n^\prime\|\le \delta_n, 1\le n \le m.
$
Given identical prior $\bm P=\{P_n\}_{n=1}^m$, the minimum failure probability $P_F^{X\star}(\bm \epsilon^X; \{\bm \rho,\bm P\})$ as a function of the tolerance $\bm \epsilon^X$
satisfies the continuity
\begin{align}
P_F^{X\star}\left(\bm \epsilon^X;\{\bm \rho,\bm P\}\right)
\ge 
P_F^{X\star}(\bm \epsilon^{X\prime}; \{\bm \rho^\prime,\bm P\})-\frac{1}{2}\bm P\cdot \bm \delta,
\label{Eq:continuity_full}
\end{align}
where we denote $\bm \delta=\{\delta_k\}_{k=1}^n$, and use the notation that $X=U$ or $X=R$ represents the two different cases. The parameter for the un-rescaled case
\be 
\bm \epsilon^{U\prime}=\bm \epsilon^U+ \bm \delta;
\ee 
While for the rescaled case 
\begin{align}
 \bm \epsilon^{R\prime}\equiv  \frac{\bm \delta+ \bm \epsilon^R \left[1-P_F^{R\star}\left(\bm \epsilon^R;\{\bm \rho,\bm P\}\right)\right]}{1-P_F^{R\star}\left(\bm \epsilon^R;\{\bm \rho,\bm P\}\right)-\frac{1}{2}\bm P\cdot \bm \delta}.
\end{align}
\end{lemma}
Here we see that the un-rescaled case has a much simpler continuity bound; in fact, by interchanging the variables, one can also show an upper bound as
\be
P_F^{U\star}\left(\bm \epsilon- \bm \delta; \{\bm \rho^\prime,\bm P\}\right)+\frac{1}{2}\bm P\cdot \bm \delta
\ge
P_F^{U\star}\left(\bm \epsilon;\{\bm \rho,\bm P\}\right).
\ee 
While in the second formalism, an upper bound will be complicated to obtain.

To prove the lemma, we will use one-norm's variational form
\begin{equation}
\|A\|=2\sup_{0\le P\le I} \mathrm{Tr}\left[P A\right],
\end{equation}
so that $|\mathrm{Tr}\left[\left(\rho_k^\prime-\rho_k\right)
\Pi\right]|\le \|\rho_k^\prime-\rho_k\|/2\le \delta_k/2$ for any POVM element $\Pi$.

Consider a measurement described by the POVM elements $\bm \Pi=\{\Pi_n\}_{n=0}^m$, below we show that its performance on the two enfeebles of states is close. First, the inconclusive probability
\be 
P_F(\bm \rho)-P_F(\bm \rho^\prime)=\sum_{k=1}^nP_k\Tr\left(\Pi_0 \left(\rho_k-\rho_k^\prime\right)\right),
\nonumber
\ee 
and therefore we have
\be 
|P_F(\bm \rho)-P_F(\bm \rho^\prime)|\le \frac{1}{2}\bm P\cdot \bm \delta.
\ee 
The conditional error probability
\be 
P_{E|k}(\bm \rho)-P_{E|k}(\bm \rho^\prime)=\Tr\left(\left(\rho_k^\prime-\rho_k\right) \Pi_k\right)+\Tr\left(\left(\rho_k^\prime-\rho_k\right) \Pi_0\right),
\nonumber
\ee 
and therefore we have
\be 
|P_{E|k}(\bm \rho)-P_{E|k}(\bm \rho^\prime)|\le \delta_k.
\ee 

Now we look at the optimum solutions.

For the first case, suppose for $\bm \rho$, the measurement achieves $ P_F^{U\star}\left(\bm \epsilon ;\{\bm \rho,\bm P\}\right)$. This means that 
\be 
P_F(\bm \rho)=P_F^{U\star}\left(\bm \epsilon ;\{\bm \rho,\bm P\}\right), P_{E|k}(\bm \rho)\le \epsilon_k.
\ee 
The performance of the same measurement on the other ensemble
\begin{align} 
&P_F(\bm \rho^\prime)\le P_F^{U\star}\left(\bm \epsilon ;\{\bm \rho,\bm P\}\right)+\frac{1}{2}\bm P\cdot \bm \delta,
\\
&P_{E|k}(\bm \rho^\prime)\le \delta_k+P_{E|k}(\bm \rho^\prime)\le \delta_k+\epsilon_k.
\end{align} 
So we have an operating point that suffices to show
\begin{align}  
P_F^{U\star}\left(\bm \epsilon+\bm \delta ;\{\bm \rho^\prime,\bm P\}\right)\le P_F^{U\star}\left(\bm \epsilon ;\{\bm \rho,\bm P\}\right)+\frac{1}{2}\bm P\cdot \bm \delta,
\end{align} 
because the minimum can only be smaller than $P_F(\bm \rho^\prime)$.

For the second case, suppose for $\bm \rho$ the measurement achieves $p_f\equiv P_F^{R\star}\left(\bm \epsilon^R;\{\bm \rho,\bm P\}\right)$. This means that
\be 
P_F(\bm \rho)=p_f, P_{E|k}(\bm \rho)\le \epsilon^R_k (1-p_f),
\ee 
and we have its performance on the other ensemble
\begin{align} 
&P_F(\bm \rho^\prime)\le p_f+\frac{1}{2}\bm P\cdot \bm \delta,
\label{proof_continuity_step1_ineq1}
\\
&P_{E|k}(\bm \rho^\prime)\le \delta_k+P_{E|k}(\bm \rho^\prime)\le \delta_k+ \epsilon^R_k (1-p_f).
\label{proof_continuity_step1_ineq2}
\end{align} 
The corresponding constraint
\be 
\frac{P_{E|k}(\bm \rho^\prime)}{1-P_F(\bm \rho^\prime)}\le \frac{\delta_k+ \epsilon^R_k (1-p_f)}{1-p_f-\frac{1}{2}\bm P\cdot \bm \delta}.
\label{xi_constraint_proof}
\ee 
Note that the above is only true for $p_f+\frac{1}{2}\bm P\cdot \bm \delta<1$, however, when $p_f+\frac{1}{2}\bm P\cdot \bm \delta\ge1$, the bound is meaningless and therefore we do not worry about that case.
Therefore the overall performance satisfies
\begin{align}
&P_F(\bm \rho^\prime)\le p_f+\frac{1}{2}\bm P\cdot \bm \delta,
\\
&\frac{P_{E|k}(\bm \rho^\prime)}{1-P_F(\bm \rho)}\le \epsilon_k^{R\prime}\equiv  \frac{\delta_k+ \epsilon^R_k (1-p_f)}{1-p_f-\frac{1}{2}\bm P\cdot \bm \delta}.
\end{align}
This operating point suffices to give an upper bound
\begin{align}
&P_F^{R\star}\left(\bm \epsilon^{R\prime};\{\bm \rho^\prime,\bm P\}\right)
\le 
P_F^{R\star}\left(\bm \epsilon^R;\{\bm \rho,\bm P\}\right)+\frac{1}{2}\bm P\cdot \bm \delta,
\end{align}

%\QZE{The following does not work out, because $P_F(\bm \rho^\prime)>P_F^{R\star}$.}
%One might also want a version where $\bm \epsilon^{R\prime}$ is not a function of $p_f$, then we can alternatively replace Ineq.~\eqref{xi_constraint_proof} with
%\begin{align} 
%\frac{P_{E|k}(\bm \rho^\prime)}{1-P_F(\bm \rho^\prime)}&\le \epsilon^R_k +\frac{\delta_k+ \epsilon^R_k \left(\frac{1}{2}\bm P\cdot \bm \delta\right)}{1-P_F(\bm \rho^\prime)}
%\end{align}
%where we used Ineqs.~\eqref{proof_continuity_step1_ineq1} and~\eqref{proof_continuity_step1_ineq2} in the first step and $P_F(\bm \rho^\prime)$.

\begin{figure}
\centering
\includegraphics[width=0.5\textwidth]{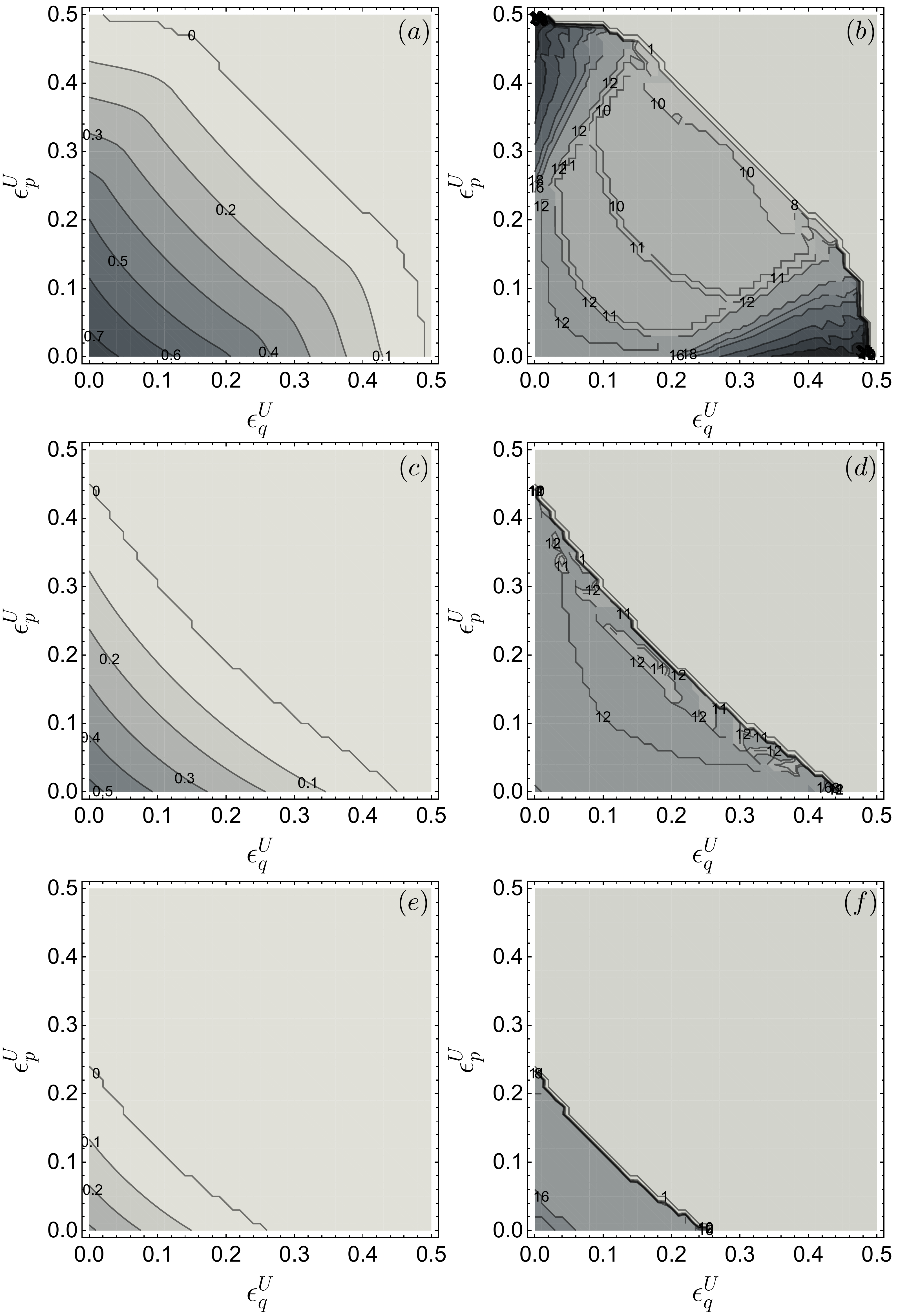}
\caption{Optimum lower bound $\max_M P_{F,u,LB,M}^U$ in Eq.~\eqref{bound_fidelity_channel} of Lemma~\ref{lemma:bound_fidelity_channel} of the main paper. The optimum lower $M$ is given in (b)(d)(f), while the inconclusive probability is in (a)(c)(e). $r_q=0.8,r_p=0.9$.
(a-b)$u=1$, (c-d)$u=2$, (e-f)$u=3$.
We numerically scan over $\epsilon_p^R,\epsilon_q^R$ to obtain both $\epsilon_p^U,\epsilon_q^U$ and $P_{F,u,LB,M}^U$; then we optimize over $M$ for each $\epsilon_p^U,\epsilon_q^U$.
\label{fig:dephasing_LB_full}
}
\end{figure}

\section{Proof of Theorem~\ref{theorem:LB}} 
\begin{proof}
In this proof, we will simplify the notation and omit the identical prior $\bm P$ in the ensemble.

First, by continuity Eq.~\eqref{Eq:continuity} in Lemma~\ref{approximate_continuity} of the main paper and applying inequality~\eqref{stretchingEQ} of the main paper on each channel, we have

\begin{align}
&P_F^{U\star}\left(\bm \epsilon;\left\{\rho_{\calE_n,u}\right\}_{n=1}^m\right)\ge
\nonumber
\\
&
P_F^{U\star}\left(\left\{\epsilon+u \Delta_{\calE_n,M}\right\}_{n=1}^m; \left\{\Lambda(\rho_{\calE_n}^{\otimes u M})\right\}_{n=1}^m\right)-\frac{1}{2}\sum_{n=1}^m u P_n\Delta_{\calE_n,M},
\\
&\ge 
P_F^{U\star}\left(\left\{\epsilon+u \Delta_{\calE_n,M}\right\}_{n=1}^m; \left\{\rho_{\calE_n}^{\otimes u M}\right\}_{n=1}^m\right)-\frac{u}{2}\sum_{n=1}^m  P_n\Delta_{\calE_n,M}.
\end{align}
In the last step, we utilized the data processing inequality~\eqref{Eq:data_processing} of the main paper.

The above inequality holds for any output states $\left\{\rho_{\calE_n,u}\right\}_{n=1}^m$, and therefore holds for an arbitrary adaptive protocol $\mathbb{P}_u$.
\end{proof}

\section{Extra numerical results}

In Fig.~\ref{fig:dephasing_LB_full}, we provide the corresponding optimum $M$ for Fig.~\ref{fig:dephasing_LB} of the main paper.

\end{document}